\documentclass[a4paper, 11pt]{article}
\usepackage[english]{babel}
\usepackage{amsmath, amssymb, amsfonts, bm}
\usepackage{mathrsfs}
\usepackage{bbm}
\usepackage{float}
\usepackage{color,graphicx,epstopdf}
\usepackage{dsfont}
\usepackage{hyperref}
\usepackage{cite} 

\newcommand*\mcap{\mathbin{\mathpalette\mcapinn\relax}}
\newcommand*\mcapinn[2]{\vcenter{\hbox{$\mathsurround=0pt
  \ifx\displaystyle#1\textstyle\else#1\fi\bigcap$}}}

\newcommand*\mcup{\mathbin{\mathpalette\mcupinn\relax}}
\newcommand*\mcupinn[2]{\vcenter{\hbox{$\mathsurround=0pt
  \ifx\displaystyle#1\textstyle\else#1\fi\bigcup$}}}

\DeclareFontFamily{OT1}{pzc}{}
\DeclareFontShape{OT1}{pzc}{m}{it}{<-> s * [1.200] pzcmi7t}{}
\DeclareMathAlphabet{\mathpzc}{OT1}{pzc}{m}{it}

\newtheorem{theorem}{Theorem}
\newtheorem{definition}{Definition}
\newtheorem{lemma}{Lemma}

\newtheorem{proposition}{Proposition}

\title{\bf Boolean Gossiping Networks
}
\date{}

\author{Bo Li, Junfeng Wu, Hongsheng Qi, Alexandre Proutiere, and Guodong Shi\thanks{B. Li  and H. Qi are with the Key Laboratory of Mathematics Mechanization and with the Key Laboratory of Systems and Control respectively, Academy of Mathematics and Systems Science, Chinese Academy of Sciences, Beijing 100190, China. Email:  libo@amss.ac.cn, qihongsh@amss.ac.cn.}
\thanks{J. Wu and A. Proutiere are with ACCESS Linnaeus Centre, KTH Royal Institute of Technology, Stockholm 10044, Sweden. Email: junfengw@kth.se,  alepro@kth.se.}
\thanks{G. Shi is with the Research School of Engineering,  The Australian National University, Canberra 0200, Australia. E-mail: guodong.shi@anu.edu.au.}}

\begin{document}

\maketitle
\begin{abstract}
This paper proposes and investigates a Boolean gossip model as a simplified but non-trivial  probabilistic Boolean network.  With positive node interactions, in view of standard theories from Markov chains, we prove  that the node states asymptotically converge to an agreement   at a binary random variable, whose  distribution is characterized  for large-scale  networks by mean-field approximation. Using combinatorial analysis, we also successfully count  the  number of communication classes of the positive Boolean network explicitly  in terms of the topology of the underlying interaction graph, where remarkably  minor variation  in  local structures can drastically change the number of network communication classes.  With general Boolean interaction rules,  emergence of  absorbing network Boolean dynamics is shown to be determined   by the network structure  with necessary and sufficient conditions established regarding when the Boolean gossip process defines  absorbing Markov chains. Particularly, it is shown that for the majority of the Boolean interaction rules, except for nine out of the total  $2^{16}-1$ possible nonempty  sets of binary Boolean functions, whether the induced chain is absorbing   has nothing to do with the topology  of the underlying interaction graph, as long as  connectivity is assumed.  These results illustrate  possibilities  of {relating dynamical} properties of Boolean networks to graphical properties of the underlying interactions.
\end{abstract}

\section{Introduction}

\subsection{Background}
A variety of random network dynamics with nodes taking  logical values  arises from biological, social, engineering, and artificial intelligence systems \cite{Kauffman1969, Hopfield-1982, Karp2000, mie09}. In the 1960s, Kauffman introduced random Boolean iteration rules over a network  \cite{Kauffman1969} to describe  proto-organisms as  randomly aggregated nets of chemical
reactions where the underlying  genes serve as a
binary (on-off) device. Inspired by neuron systems, the so-called Hopfield networks \cite{Hopfield-1982} provided a way of realizing  collective computation intelligence, where nodes having binary values behave as artificial neurons by a weighted majority voting  via random or deterministic updating. Rumors spreading  over a social network \cite{Karp2000} and virus scattering  over a computer network \cite{mie09} can be modeled as epidemic processes with binary nodes states indicating   whether a peer has  received a  rumor, or whether a computer has been infected by a type of virus.

 Boolean dynamical networks, consisting  of a finite set of nodes and a set of deterministic or {random} Boolean interaction rules among the nodes,  are natural and primary tools for the modeling of the above node dynamics with logical values. The study of Boolean  networks  received considerable attention for aspects ranging from steady-state behaviors and  input-output relations  to  limit cycle attractors and model reduction, e.g., \cite{Probabilistic-Boolean-Network,Brun2005,Shmulevich2003,Cheng2009,Cheng-Qi2010,Chaves2013,Chaves2014,TSP-2010-Ral,TSP-2010,m1,m2}. It has been well understood that deterministic Boolean rules are essentially linear in the state space \cite{Cheng2009,Cheng-Qi2010}, while probabilistic Boolean networks are merely standard Markov chains \cite{TSP-2010-Ral,TSP-2010,Chaves2013,Chaves2014,m1,m2}.  There however exist   fundamental challenges  in establishing  explicit and precise theoretical results due to computation complexity barriers \cite{np-hard} and the lack of analytical  tools.

In this paper, we propose and study a randomized Boolean gossip process, where Boolean nodes pairwise  meet over an underlying graph in a random manner  at each time step, and then  the two interacting  nodes update their states by random logical rules in a prescribed set of Boolean operations.

\begin{figure*}[ht]
\centering
 \includegraphics[width=17.5cm]{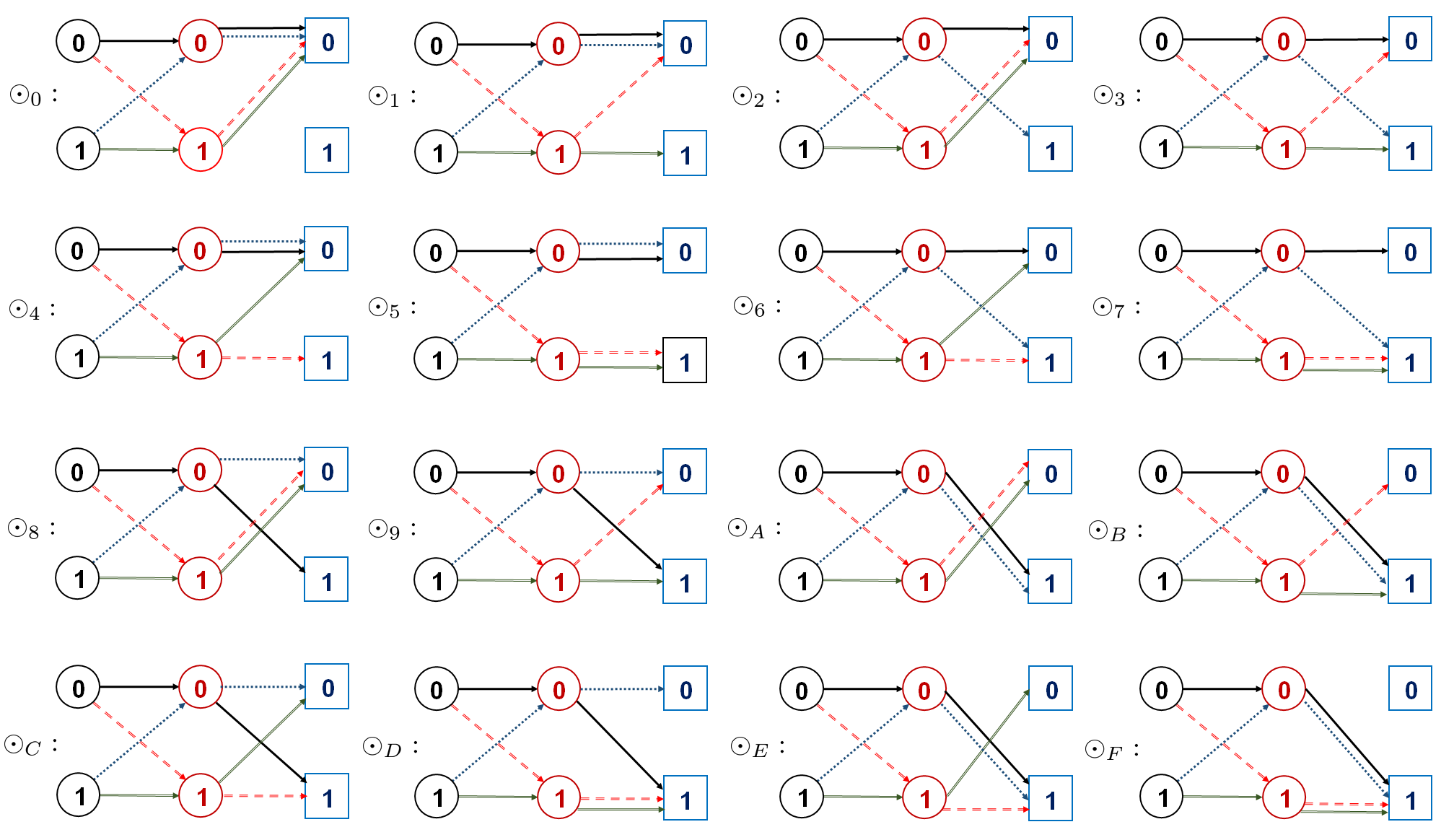}
 \caption{The 16 Binary operators mapping from $\{0,1\}^2$ to $\{0,1\}$. Each diagram visualizes a Boolean mapping: the first column  represents values of the first argument (in black); the second column represents  values of the second argument (in red); the third column (in blue) represents the outcome of the operation following the direction of the same type of lines.   For example, the first diagram reads as $0\odot_0 0 =0$, $0\odot_01=0$, $1\odot_0 0 =0$, $1\odot_0 1 =0$.} \label{fig:booleanfunctions}
 \end{figure*}
\subsection{The Model}

We consider $n$ nodes indexed {by} the set $\mathrm{V}=\{1,\dots,n\}$.
The underlying interaction structure of the network is modeled by  an undirected  graph $
\mathrm{G}=(\mathrm{V},\mathrm{E}),
$ where $\mathrm{E}$ is  the edge set with each entry being an unordered pair of two distinct nodes in $\mathrm{V}$. The set $\mathrm{N}_i=\{j: \{i,j\}\in \mathrm{E}\}$ represents  the neighbourhood of node $i$. Throughout our paper we assume that the graph $\mathrm{G}=(\mathrm{V}, \mathrm{E})$ is connected.

 Time is slotted at $t=0,1,\dots$.  Node interactions follow a random gossip process \cite{boy06}, where independently at each time $t\geq0$,
 a pair of nodes $i$ and $j$  with  $\{i,j\}\in\mathrm{E}$ is randomly selected over the graph. Each node $i$ holds a binary value from the set $\{0,1\}$ at each time $t$, denoted $x_i(t)$.
Note that, there are a total of $16$ Boolean functions with two arguments  mapping from $\{0,1\}^2$ to $\{0,1\}$. Using  hexadecimal numbers, we index  these functions in the set (see Fig. \ref{fig:booleanfunctions}) $$
\mathsf{H}:=\{\odot_0,\dots,\odot_9,\odot_A,\dots,\odot_{F}\},
$$
 where\footnote{These Boolean functions have their respective names, for which we refer to \cite{Simpson1987}.} each $\odot_k$ specifies a binary Boolean function in the way that $a \odot_k b$ is the value of the function  with arguments $(a,b)$.
Let $\mathsf{C}\neq \emptyset$ be a subset of $\mathsf{H}$ specifying potential node interaction rules along the edges. Let $q:=|\mathsf{C}|$ be the cardinality of the set $\mathsf{C}$.  We index the elements in $\mathsf{C}$ by
$$
\odot_{\mathsf{C}_1}, \dots, \odot_{\mathsf{C}_q}.
$$

 {Suppose the node pair $\{i,j\}$ is selected at time $t$.}
 Introduce  $p_1,\dots,p_q>0$ satisfying $\sum_{k=1}^q p_k=1$. Independent with time and pair selections,  the evolution of the $x_m(t)$ is determined by
 \begin{equation}\label{probabilistic-pair}
\begin{cases}
		x_i(t+1)= x_i(t)  \odot_{\mathsf{C}_k} x_j(t), & \text{with prob. $p_k,k=1,\dots,q$;}\\
		x_j(t+1)= x_j(t)  \odot_{\mathsf{C}_l} x_i(t), & \text{with prob. $p_l,l=1,\dots,q$;}\\
x_m(t+1)= x_m(t), & \text{$m\notin \{i,j\}$},
	\end{cases}
\end{equation}
where the updates of nodes $i$ and $j$ are independent with each other.

\subsection{Induced Markov Chain}
Let $X_t=(x_1(t),\dots,x_n(t)), t=0,1,\dots$ be the random process driven by the { gossip algorithm and the Boolean rules} (\ref{probabilistic-pair}).  This random process $X_t,t\geq0$ defines a $2^n$-state Markov chain $\mathcal{M}_\mathrm{G}(\mathsf{C})=(\mathbf{S}_n, P)$, where
$$
\mathbf{S}_n=\big\{[s_1 \dots s_n]:\ s_i\in\{0,1\}, i\in \mathrm{V} \big\}
$$
is the state space, and $P$ is the state transition matrix. Then the state transition matrix $P$ is given by
$$P=\big[P_{[s_1 \dots  s_n][q_1 \dots  q_n]}\big]\in \mathbb{R}^{2^n \times 2^n}$$
with its rows and columns indexed by the elements in $\mathbf{S}_n$, i.e.,
$$
P_{[s_1 \dots  s_n][q_1 \dots  q_n]}:=\mathbb{P}\Big (X_{t+1}=[q_1 \dots  q_n]  \Big|  X_t=[s_1 \dots  s_n] \Big).
$$

\subsection{Related Work}

The proposed randomized Boolean gossip model apparently cover the classical gossip process  \cite{boy06,dar10,Doerr2012,sha08} as a special case.  The process (\ref{probabilistic-pair}) is also a special case of the probabilistic Boolean network model \cite{Probabilistic-Boolean-Network,Shmulevich2003}, where random Boolean interactions are posed pairwise. Therefore  conceptually the model (\ref{probabilistic-pair}) under consideration  can certainly  be placed into the studies of general probabilistic Boolean networks, e.g.,  \cite{Brun2005,TSP-2010-Ral,TSP-2010}. Since the node interaction rules can be an arbitrary set of Boolean functions, this Boolean gossip model is a useful approximation or generalization to existing characterizations to gene regulation \cite{Kauffman1969}, social opinion evolution \cite{Karp2000}, and virus spreading \cite{mie09}.

\medskip

\noindent {\em Gene Regulation}. The evolution of gene expressions can be naturally described
as a dynamical system where the two quantized levels,
ON and OFF, are represented by logic states 1 and 0,
respectively.  Each gene normally would
only interact with a small number of  neighbouring  genes\footnote{Such number is
two or three in Kauffman's original  proposal \cite{Kauffman1969}.}. Therefore,
the proposed Boolean gossip network model at least serves as  a good approximation for   gene regulator
networks, where a pair of genes interact  at any given time and the Boolean function rules $\mathsf{C}$ describe  random  outcomes of the interactions.

\medskip

\noindent {\em Social Voting}. Social peers  hold binary opinions  for certain political or economical issues, where $1$ represents a supportive opinion and $0$ represents a non-supportive one. Peers meet with each other in pairs randomly and exchange their opinions.  The two peers independently decide their {opinions} after the meeting;  the Boolean function rules $\mathsf{C}$ describe  how they might revise their opinions.

\medskip

\noindent {\em Virus Spreading}. Virus  spreading  across   a computer network can be modeled as a Boolean network, where $0$ and $1$  represent infected and  healthy  computers, respectively \cite{mie09}. The proposed Boolean gossip process may characterize  more possibilities for two computers during an interaction:  two  computers, infected or not,  are both infected  ($\odot_1$);  two  computers, infected or not,  are both cured ($\odot_F$), etc.

\medskip

The graphical nature of the model  (\ref{probabilistic-pair})    makes it possible to go beyond these existing work \cite{Brun2005,TSP-2010-Ral,TSP-2010} for  more direct and explicit  results.  Additionally, majority Boolean dynamics \cite{AAP-2011} and asynchronous broadcast gossiping \cite{Amini-2013} are related to the model (\ref{probabilistic-pair}) in the way that they describe  Boolean interactions between one node and all its neighbors at a given time instant, in contrast to the gossip interaction rule  which happens between one node and one of its selected neighbors.
\subsection{Contributions and Paper Organization}
 The proposed random Boolean gossip model  is fully determined by the underlying graph $\mathrm{G}$ and the Boolean interaction set $\mathsf{C}$. Classical (deterministic or probabilistic) Boolean networks also have graphical characterization \cite{Probabilistic-Boolean-Network} where a link appears  if the state of the end nodes depend on each other in the Boolean updating rules. To the best of our knowledge, few results have been obtained regarding how the structure of the interaction graph influences detailed network state evolution  in the study of Boolean networks.

 First of all, we study  a special network where the Boolean interaction rules in the set $\mathsf{C}$ do not involve the negation, which is termed positive Boolean networks. Using standard theories from Markov chains, we show that the network nodes asymptotically converge to a consensus  represented by a binary random variable, whose  distribution is studied  for large-scale  networks in light of mean-field approximation methods. Moreover, by combinatorial analysis the  number of communication classes of positive Boolean networks is fully characterized with respect to the structure  of the underlying interaction graph $\mathrm{G}$, where surprisingly  local cyclic   structures can drastically change the number of communication classes of the  entire   network.

  Next, we move to general Boolean interaction rules and study the relation between emergence of  absorbing network Boolean dynamics and the network structure.  Necessary and sufficient conditions are provided for the induced Markov process $\mathcal{M}_\mathrm{G}(\mathsf{C})$ to be an absorbing chain. Interestingly,  for the majority of the Boolean interaction rules, except for nine of  the $2^{16}-1$ possible nonempty  sets of binary Boolean functions, whether the induced chain is absorbing does not rely on   the network topology as long as the underlying graph is connected; for the remaining  nine sets of binary Boolean functions,  absorbing property of the induced chain is fully determined by whether  the underlying graph $\mathrm{G}$  contains an odd cycle.

The remainder of this paper is organized follows. Section \ref{sec:positive} investigates positive Boolean dynamics in terms of steady-state distribution and communication classes. Section  \ref{sec:general} further studies general Boolean dynamics with a focus on how the interaction graph determines absorbing Markov chains along  the random  Boolean dynamics. Finally Section \ref{sec:conclusion} concludes the paper with a few remarks.

\section{Positive Boolean Gossiping}\label{sec:positive}
In this section, we consider a special case where the Boolean interaction rules in the set $\mathsf{C}$ do not involve the negation $\neg$. Note that conventionally ``$\wedge$" represents Boolean ``AND" operation, while ``$\vee$" represents Boolean ``OR" operation.  We term such types of Boolean interaction as {\it positive} Boolean dynamics, and define  $$
\mathsf{C}_{\rm pst} =\{ \vee, \wedge\}
$$
as the set of positive Boolean functions. Let us denote $\odot_{\mathsf{C}_1}=\vee$ and $\odot_{\mathsf{C}_2}=\wedge$. Let $p_*=p_1$ be the probability for operation ``$\vee$" in the dynamics (\ref{probabilistic-pair}).

\subsection{State Convergence}

Recall that a state in a Markov chain  is called {\it absorbing} if it is impossible to leave this state \cite{Gri12}. A Markov chain is called absorbing  if
it contains  at least one absorbing state and  it is possible to go from any state to at least one absorbing state in a finite number of steps. In an absorbing Markov chain, the non-absorbing states are  called {\it transient}.

It is not hard to find that the Markov chain $\mathcal{M}_\mathrm{G}(\mathsf{C}_{\rm pst})$ is an absorbing chain with
$[0\dots 0]$ and $[1\dots 1]$ being the two absorbing states. Let
$I_k$ denote the $k$-by-$k$ identity matrix for any integer $k$.
The state transition matrix
$P$ therefore will have the form
\[
P=
\left[
\begin{array}{c|c}
I_2 & 0 \\
\hline
R & Q
\end{array}
\right],
\]
where the $I_2$ block  corresponds to the two absorbing states $[0\dots 0]
$ and $[1\dots 1]$, $R$ is a $(2^n-2)\times 2$ matrix describing transition from the $2^n-2$ transient states to the two absorbing states, and $Q$ is a $(2^n-2)\times (2^n-2)$ matrix describing the transition between the transient states.

Note that following the definition of $P$, the rows of the matrix $(I_{2^n-2} -Q)^{-1} R$ are indexed by the entries in $\mathbf{S}_n \setminus \{[0\dots 0],[1\dots 1]\}$, and the columns are indexed by $[0\dots 0]$ and $[1\dots 1]$. Let $\big[(I_{2^n-2} -Q)^{-1} R\big]_{X_0[1\dots 1]}$ be the $X_0$-$[1\dots 1]$ entry of the matrix $(I_{2^n-2} -Q)^{-1} R$.
 We can conclude the following result from standard theories for absorbing Markov chains (see Theorem 11.6, pp. 420, \cite{Gri12}).

 \medskip

\begin{proposition}\label{proposition1}
Let $X_0=X(0)\in\mathbf{S}_n \setminus \{[0\dots 0],[1\dots 1]\}$. There exists a  Bernoulli random variable $x_\ast$ such that
$$
\mathbb{P}\big(\lim_{t\rightarrow \infty} x_i(t) = x_\ast, \mbox{for all}\ i\in\mathrm{V}\big)=1.
$$
  The limit $x_\ast$ satisfies
 $$
 \mathbb{E}\{x_\ast\}=\big[(I_{2^n-2} -Q)^{-1} R\big]_{X_0[1\dots 1]}.
 $$

\end{proposition}

\subsection{Communication Classes}
We continue to investigate  the communication classes of $\mathcal{M}_\mathrm{G}(\mathsf{C}_{\rm pst})$. Recall that a state $[s_1 \dots  s_n]$ is said to be {\it accessible} from state $[q_1 \dots  q_n]$  if there is a nonnegative integer $t$ such that $\mathbb{P}\big(X_t=[s_1 \dots  s_n]\ \big|\ X_0=[q_1 \dots  q_n]\big)>0$. It is termed that $[s_1 \dots  s_n]$ {\it communicates} with state $[q_1 \dots  q_n]$ if  $[s_1 \dots  s_n]$ and  $[q_1 \dots  q_n]$  are accessible from  each other~\cite{Gri12}.
This communication relationship forms an equivalence relation among the states  in $\mathbf{S}_n$. The equivalence classes of this relation are called {\it  communication classes} of the chain $\mathcal{M}_\mathrm{G}(\mathsf{C}_{\rm pst})$. The number of communication classes of $\mathcal{M}_\mathrm{G}(\mathsf{C}_{\rm pst})$ is denoted as $\chi_{_{\mathsf{C}_{\rm pst}}}(\mathrm{G})$. The following theorem provides a full characterization to $\chi_{_{\mathsf{C}_{\rm pst}}}(\mathrm{G})$.

\medskip

\begin{theorem}\label{thm:class}  There hold
\begin{itemize}
\item[(i)] $\chi_{_{\mathsf{C}_{\rm pst}}}(\mathrm{G})=2n$, if $\mathrm{G}$ is a line graph;
\item[(ii)] $\chi_{_{\mathsf{C}_{\rm pst}}}(\mathrm{G})=m+3$, if $\mathrm{G}$ is a cycle graph with $n=2m$; $\chi_{_{\mathsf{C}_{\rm pst}}}(\mathrm{G})=m+2$, if $\mathrm{G}$ is a cycle graph with $n=2m+1$;
\item[(iii)]  $\chi_{_{\mathsf{C}_{\rm pst}}}(\mathrm{G})=5$, if $\mathrm{G}$ is neither a line nor a cycle, and contains no odd cycle;
\item[(iv)] $\chi_{_{\mathsf{C}_{\rm pst}}}(\mathrm{G})=3$, if $\mathrm{G}$ is not a cycle graph but contains an odd cycle.
\end{itemize}
\end{theorem}

\medskip

Established  by  constructive  proofs  that can overcome  the   fundamental  computational obstacle in analyzing large-scale Boolean networks, Theorem \ref{thm:class} reveals  how local  structures can drastically change the number of communication classes as a global property of networks. The detailed proof of Theorem \ref{thm:class} has been put in the Appendix. Below we present a few examples illustrating the statements of Theorem \ref{thm:class}.

\medskip

\begin{figure}[h]
\centering
 \includegraphics[width=6cm]{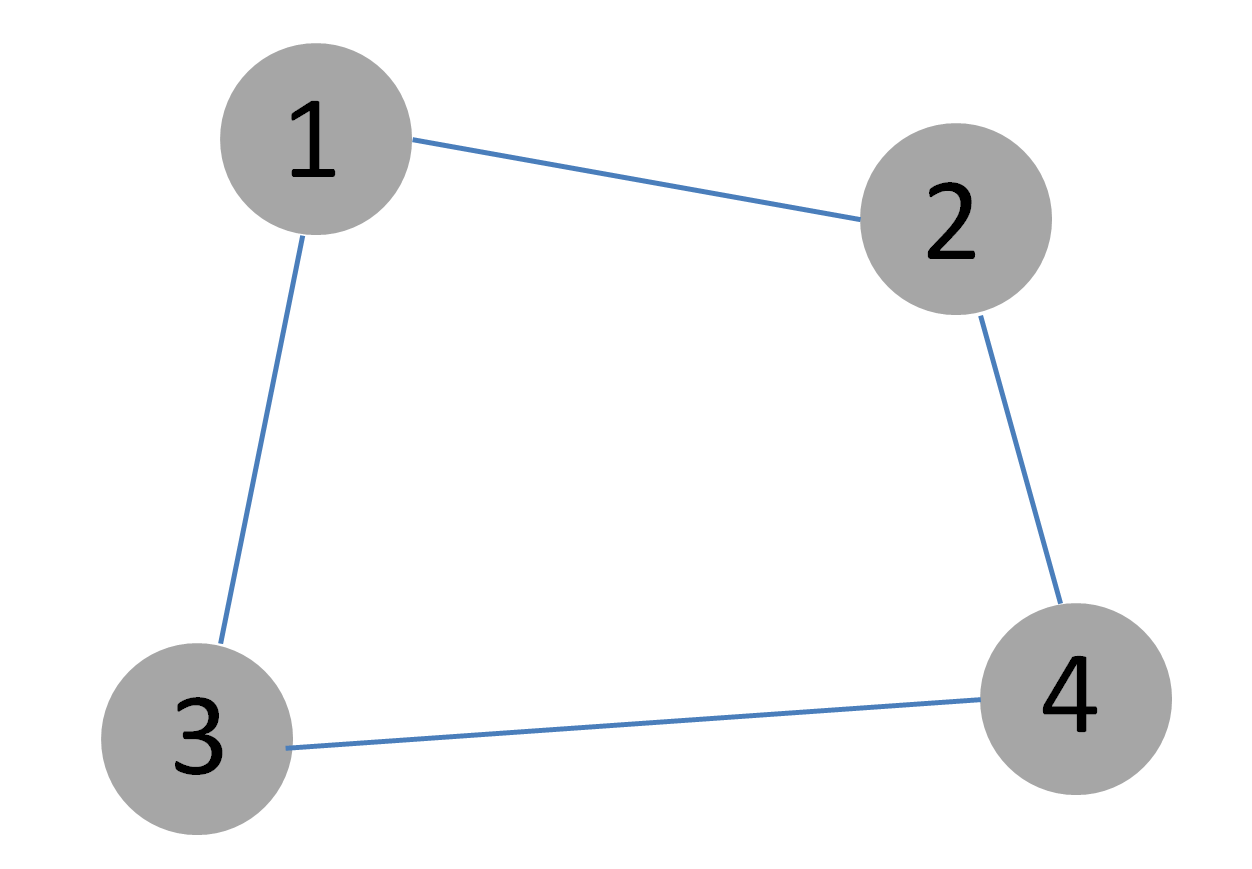}
 \caption{A four-node cycle graph.  }\label{fig:graph1}
 \end{figure}

\begin{figure}[h]
\centering
 \includegraphics[width=10.6cm]{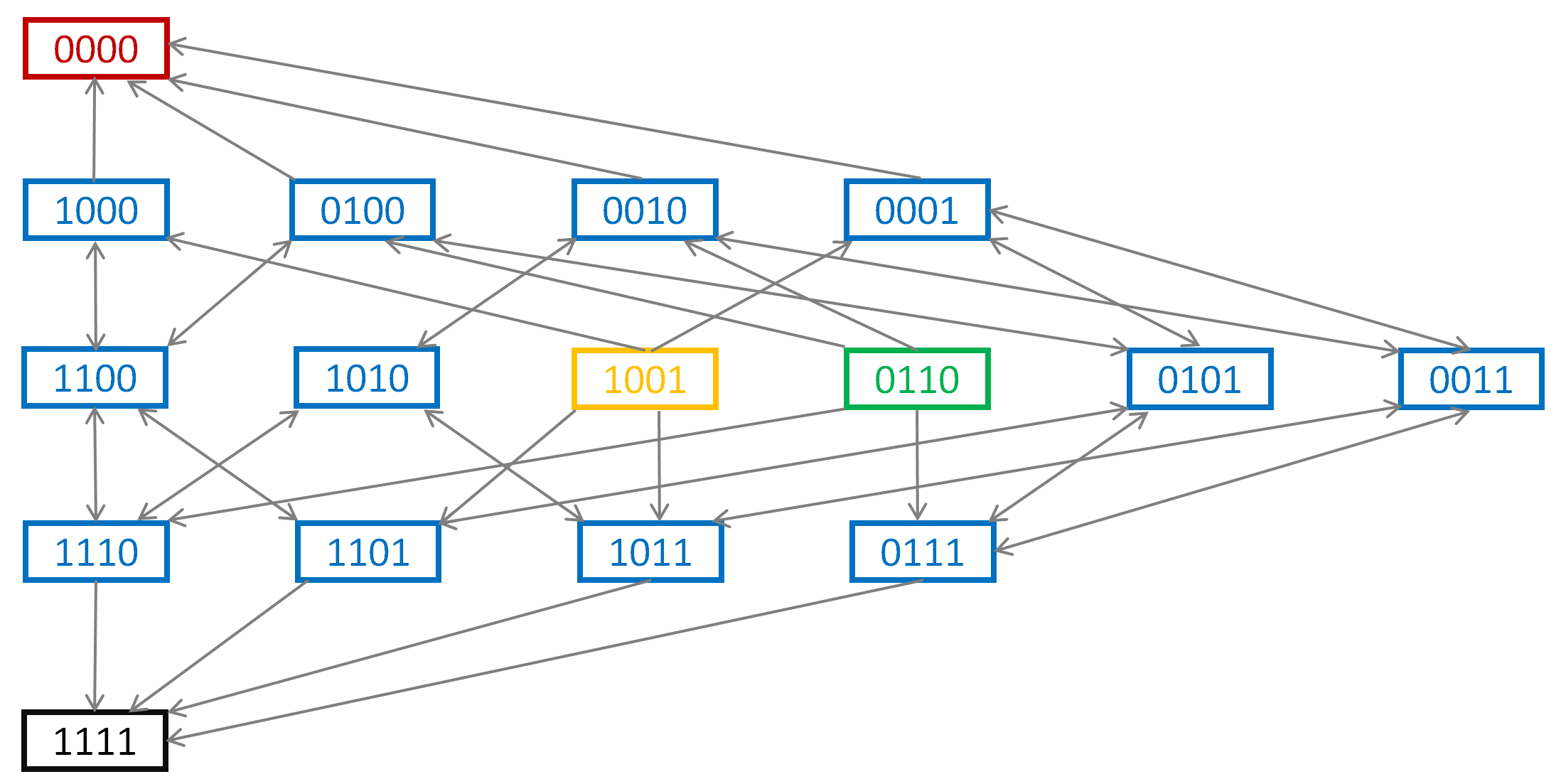}
 \caption{Full state transitions of the induced Markov chain by the positive Boolean gossip process $\mathsf{C}_{\rm pst} =\{ \vee, \wedge\}
$ over the four-node cycle graph   as shown in Figure \ref{fig:graph1}. States within the same communication class are marked with the same color.  }\label{fig:commclass1}
 \end{figure}

{\em \noindent Example 1.} Let the underlying graph $\mathrm{G}$ be the four-node cycle graph as displayed in Figure \ref{fig:graph1}. With {the} positive Boolean rules $\mathsf{C}_{\rm pst}$, the state transition map of the induced Markov chain is illustrated  in Figure \ref{fig:commclass1}. Clearly the chain has $5$ communication classes, consistent with Theorem \ref{thm:class}.

\medskip

\medskip

\begin{figure}[h]
\centering
 \includegraphics[width=5cm]{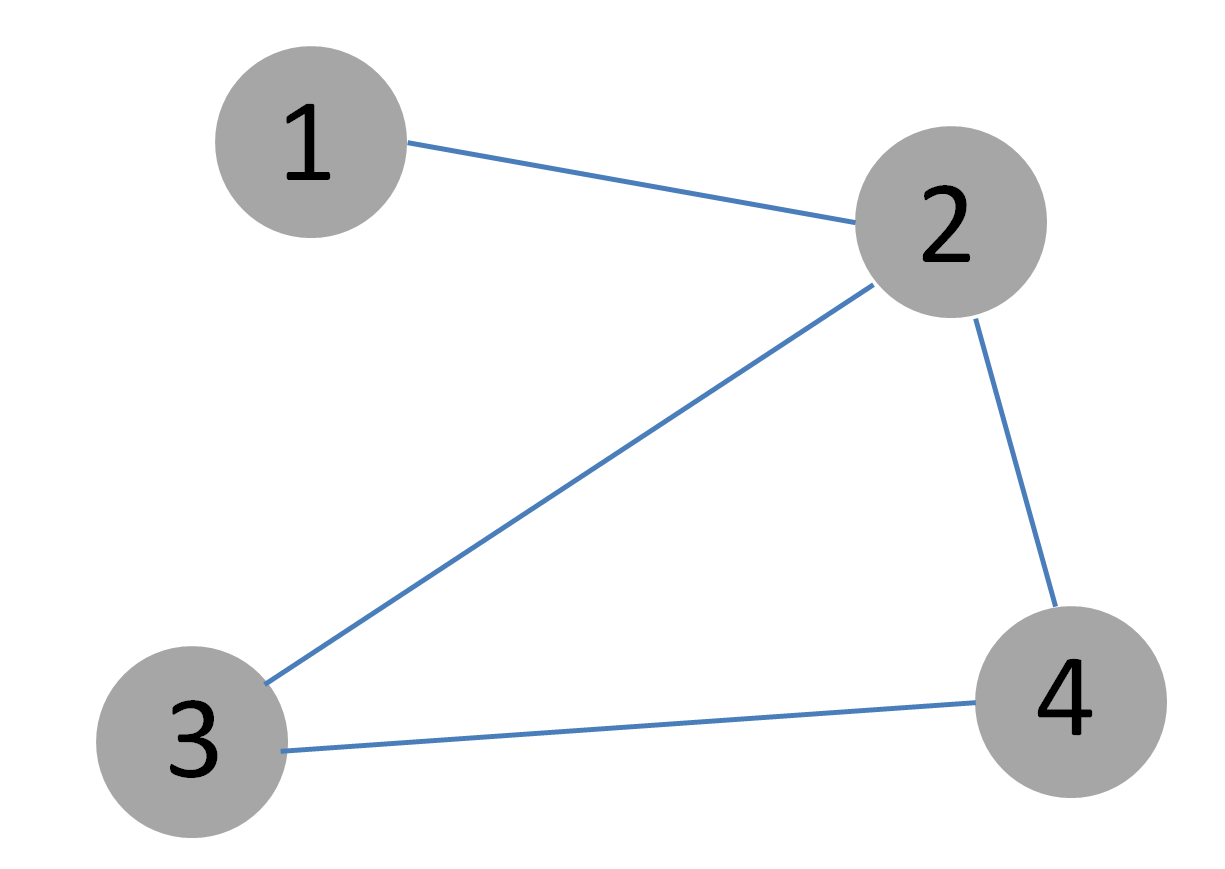}
 \caption{A four-node graph consisting of a three-node cycle subgraph.  }\label{fig:graph2}
 \end{figure}

\begin{figure}[h]
\centering
 \includegraphics[width=10cm]{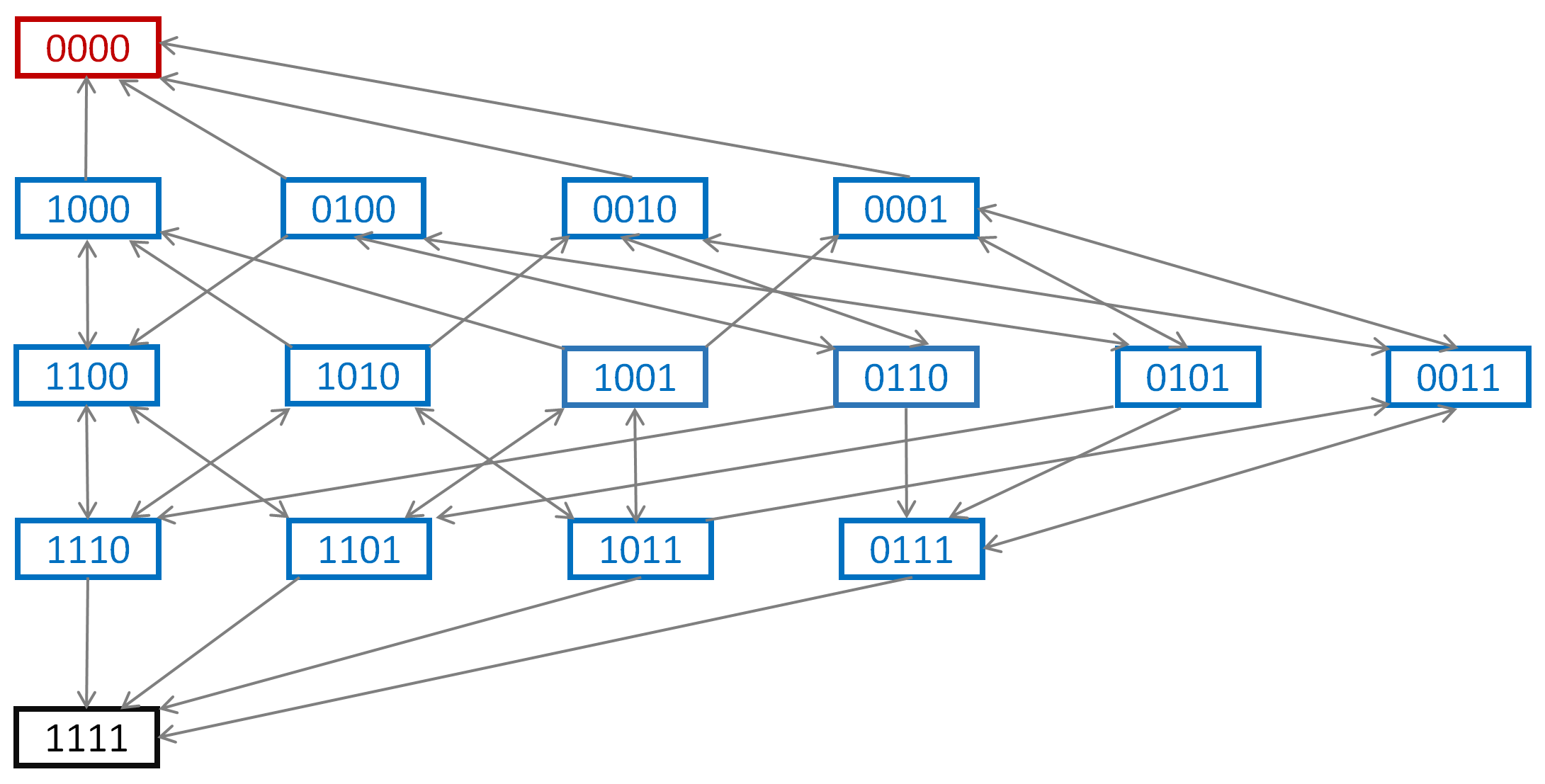}
 \caption{Full state transitions of the induced Markov chain by the positive Boolean gossip process  $\mathsf{C}_{\rm pst} =\{ \vee, \wedge\}
$ over the four-node  graph as shown in Figure \ref{fig:graph2}. States within the same communication class are marked with the same color.  }\label{fig:commclass2}
 \end{figure}

{\em \noindent Example 2.} Let the underlying graph $\mathrm{G}$ be the four-node  graph containing a three-node cycle subgraph as displayed in Figure \ref{fig:graph2}. With {the} positive Boolean rules $\mathsf{C}_{\rm pst}$, the state transition map of the induced Markov chain is illustrated  in Figure  \ref{fig:commclass2}.  In this case the chain has $3$ communication classes, again verifying  Theorem~\ref{thm:class}.

\medskip

\subsection{Continuous-Time Approximation}

It has been clear from Proposition \ref{proposition1} that starting from $X_0\in \mathbf{S}_n\setminus \{[0\dots 0],[1\dots 1]\}$, the limit of the node states is fully characterized by $\big[(I_{2^n-2} -Q)^{-1} R\big]_{X_0[1\dots 1]}$. However, computing the exact value or even obtaining an approximation for  the matrix $(I_{2^n-2} -Q)^{-1} R$ is difficult for large networks due to the exponentially increasing dimension of the matrix.
In this subsection, using mean-field method \cite{kep91,mie09}, we construct a continuous-time differential equation to approximate the behavior of $X(t)$ for large scale networks (see \cite{survey-appximation} for a detailed survey on differential equation approximations for Markov chains). To this end, we assume that the $x_i(0)$ are i.i.d Bernoulli random variables.

\subsubsection{Complete Graph}

Define
$$
\delta(t)=\sum_{i=1}^n x_i(t)/n
$$
as the proportion of nodes that take value 1 at time $t$. Assume the underlying network forms a complete graph.  Let the edges be selected uniformly at random at each time step.  Denote $\bm{\delta}(t)$ as the expected value of $\delta(t)$, i.e., $\bm{\delta}(t) = \mathbb{E}\{\delta(t)\}$.

The density $\delta(t)$ evolves by the following rules:
\begin{itemize}
\item Let the two nodes in the selected pair $\{i, j\}$ hold different values. When $n$ is large, and the graph is complete, this happens with an approximate probability $2\delta(t)(1-\delta(t))$. The value $\delta(t)$ will increase by ${1}/{n}$ if the two selected nodes both use ``$\vee$" operations to update their values, an event with probability ${p_*}^2$. The value $\delta(t)$ will decrease by ${1}/{n}$ if the two selected nodes both apply ``$\wedge$" operations, an event with probability ${(1-p_*)}^2$.
\item    For all other cases, $\delta(t)$ is unchanged.
\end{itemize}

As a result, we conclude that
\begin{align}
&\mathbb{E}\{\delta(t+1)-\delta(t)|\delta(t)\}\approx\frac{1}{n}{p_*}^2\cdot 2\delta(t)(1-\delta(t))\nonumber\\
&\ \  - \frac{1}{n}{(1-p_*)}^2\cdot2\delta(t)(1-\delta(t)).
\end{align}
For a complete graph with $n$ nodes, $\mathbb{V}\{\delta(t)\}=\mathbb{E}\{\delta^2(t)\}-\mathbb{E}^2\{\delta(t)\}$ can be considered very small for large $n$. We further have
\begin{align}\label{eqn:expectdiff}
&\bm{\delta}(t+1)-\bm{\delta}(t) \nonumber\\
&\approx \frac{1}{n}{p_*}^2\cdot 2\bm{\delta}(t)(1-\bm{\delta}(t)) - \frac{1}{n}{(1-p_*)}^2\cdot2\bm{\delta}(t)(1-\bm{\delta}(t)).
\end{align}
Define $s = {t}/{n}$ and $\tilde{\bm{\delta}}(s) =\bm{\delta}(ns)=\bm{\delta}(t)$. Then, (\ref{eqn:expectdiff}) can be written as
\begin{align}\label{eqn:expectrescale}
&\tilde{\bm{\delta}}(s+ {1}/{n}) - \tilde{\bm{\delta}}(s) \nonumber\\
&\approx \frac{1}{n}{p_*}^2\cdot 2\tilde{\bm{\delta}}(s)(1-\tilde{\bm{\delta}}(s)) - \frac{1}{n}{(1-p_*)}^2\cdot2\tilde{\bm{\delta}}(s)(1-\tilde{\bm{\delta}}(s))
\end{align}
We can therefore approximate (\ref{eqn:expectrescale}) for large $n$ by the following differential equation
\begin{equation}
\frac{d}{ds}\tilde{\bm{\delta}}(s)={p_*}^2 \cdot 2\tilde{\bm{\delta}}(s)(1-\tilde{\bm{\delta}}(s))-{(1-p_*)}^2\cdot2\tilde{\bm{\delta}}(s)(1-\tilde{\bm{\delta}}(s)),
\end{equation}
whose solution reads analytically as
\begin{equation}
\tilde{\bm{\delta}}(s)=\frac{\tilde{\bm{\delta}}(0)}{(1-\tilde{\bm{\delta}}(0))e^{2(1-2p_*)s}+\tilde{\bm{\delta}}(0)}.
\end{equation}
Here $\tilde{\bm{\delta}}(0) = \bm{\delta}(0) = \bm{\delta}_0$ is the mean of the i.i.d Bernoulli random variables $x_i(0)$.
Consequently, we establish the following approximate equation for $\bm{\delta}(t)$:
\begin{equation}\label{eqn:numsolu}
\bm{\delta}(t) = \frac{\bm{\delta}_0}{(1-\bm{\delta}_0)e^{2(1-2p_*)t/n}+\bm{\delta}_0}.
\end{equation}
From (\ref{eqn:numsolu}), the following holds.

{\noindent\em Conclusion}.  Assume $\mathrm{G}$ is a complete graph. For large $n$, $\bm{\delta}(t)$ approaches zero  when  $p_*<{1}/{2}$,  and $\bm{\delta}(t)$ approaches one when $p_*>{1}/{2}$, as time tends to infinity.

To verify this conclusion, we give some numerical results.

\medskip

\begin{figure}[h]
\centering
 \includegraphics[width=10.7cm]{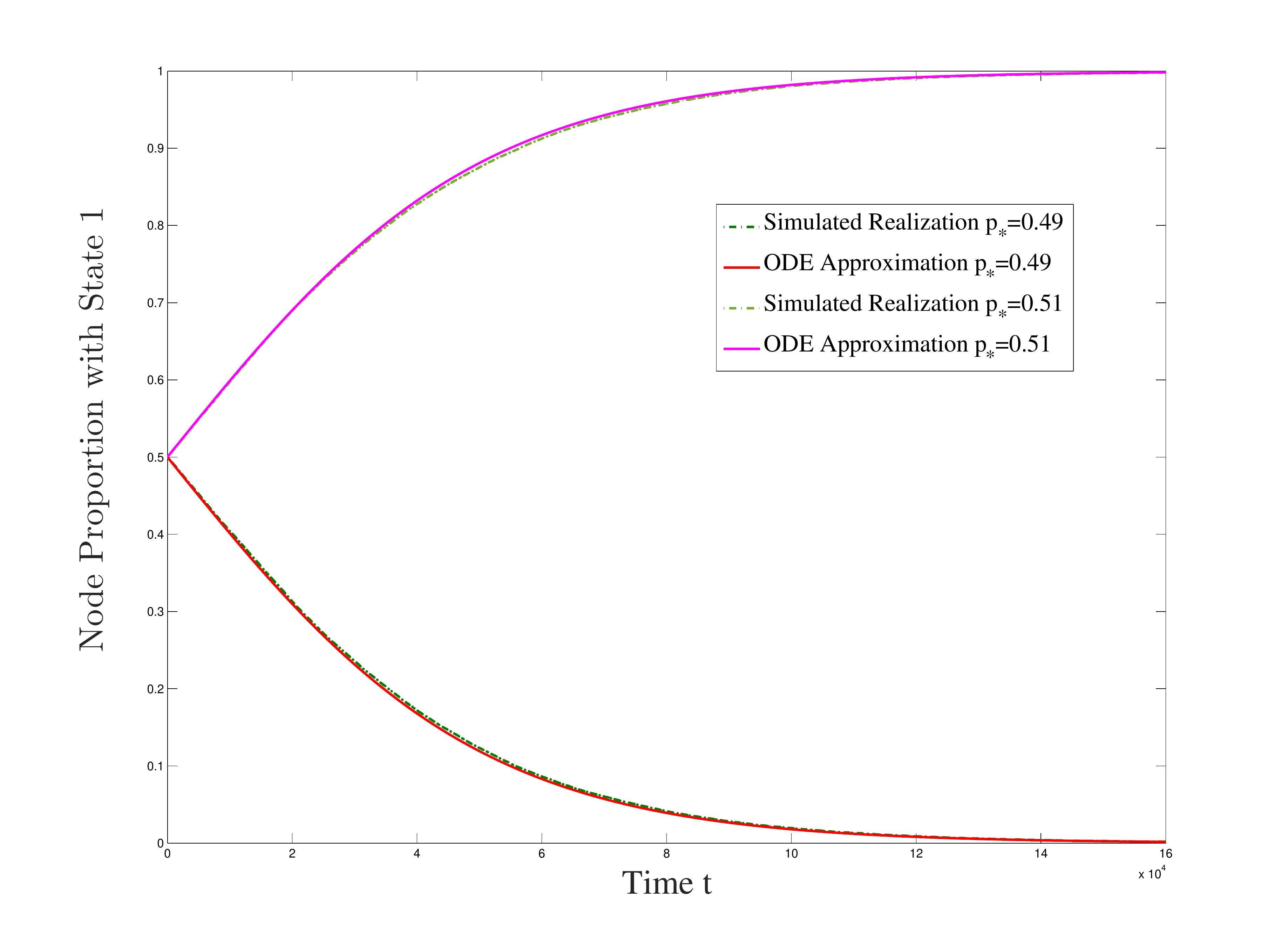}
 \caption{A complete graph with $1000$ nodes is considered. The solid lines are the approximate solution given by (\ref{eqn:numsolu});  the dashed lines are drawn according to the simulated realization of the algorithm (\ref{probabilistic-pair}). The continuous-time approximations match the numerical realizations rather precisely.}\label{fig:example1}
 \end{figure}

{\em \noindent Example 3.} Consider a complete graph with $n=1000$ nodes. Fix $\bm{\delta}_0=0.5$, and we randomly distribute the values of nodes according to $\bm{\delta}_0=0.5$.  For $p=0.49$ and $0.51$, we let the nodes update their values {randomly} according to (\ref{probabilistic-pair}), respectively.  Each experiment is carried out over $T=160000$ time steps, repeated for $2000$ rounds. The average of the resulting $2000$ sample paths approximately give the density of nodes with value one for every $t$. We compare the numerical simulation with the approximate solution given by (\ref{eqn:numsolu}). Figure \ref{fig:example1}  shows that (\ref{eqn:numsolu}) approximates the real process (\ref{probabilistic-pair}) remarkably well.

\medskip

\subsubsection{Regular Graph}
A regular graph is a graph where nodes have equal degrees. Suppose node $i$ is selected to initialize a gossip interaction at time $t$. Because $i$ is uniformly selected from $\mathrm{V}$, the probability that the selected node $i$ is at state $1$ is $\delta(t)$. If $\mathrm{G}$ is a regular graph with a random nature\footnote{This is to say, the distribution of the links should appear somehow  independently being  close to  the concentration  of random regular graphs. The approximation can be quite inaccurate for graphs like  lattices.  } and  high node degrees where $|N_i|=\mathcal{O}(n)$, the distribution of the random variable
\[
\frac{\sum_{j\in \mathrm{N}_i}x_j(t) }{ |\mathrm{N}_i|}
\]
will tend to have a similar distribution with
\[
\frac{\sum_{j\neq i}x_j(t)} { n-1},
\]
which is approximately a Bernoulli random variable with mean $\delta(t)$. Therefore, $\delta(t)$ evolves following similar rule as complete graphs, and the differential equation  (\ref{eqn:numsolu}) will continue to be a good approximation for high-degree regular graphs.

\begin{figure}[h]
\centering
 \includegraphics[width=10.7cm]{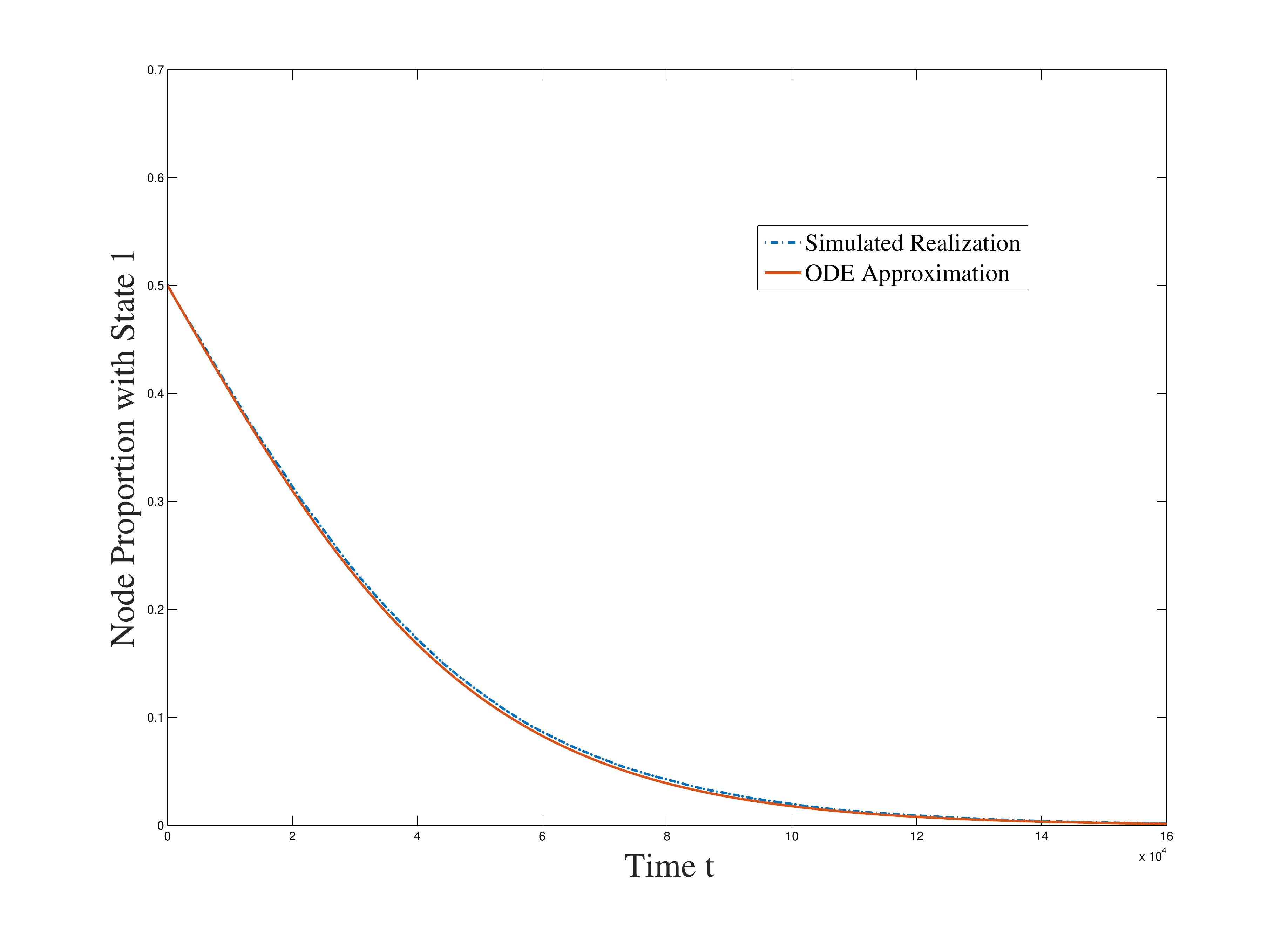}
 \caption{A regular graph with $1000$ nodes is considered where node degree is $500$ and $p_*=0.49$. The solid line is the approximate solution given by (\ref{eqn:numsolu}) and the dashed line is drawn according to numerical simulation. We see that (\ref{eqn:numsolu}) continues to be a good approximation of (\ref{probabilistic-pair}).}\label{fig:example2}
\end{figure}

\medskip
{\em \noindent Example 4.} Consider a regular graph of degree $500$ with $n=1000$ nodes. We select $p=0.49$ and $\bm{\delta}_0=0.5$. Again each experiment is carried out over $T=160000$ time steps, repeated for $2000$ rounds. The average of the resulting $2000$ sample paths allows us to obtain the approximate  density of nodes with value $1$ for all $t$. Figure \ref{fig:example2}   shows that (\ref{eqn:numsolu}) continues to provide an acceptable approximation of the real process (\ref{probabilistic-pair}).

\medskip

%
%

\section{General Boolean Dynamics}\label{sec:general}
In this section, we discuss the evolution of (\ref{probabilistic-pair}) under general Boolean interaction set $\mathsf{C}\in 2^{\mathsf{H}}$, where $2^{\mathsf{H}}$ denotes the set containing all subsets of $\mathsf{H}$. {We are interested in how the induced chain $\mathcal{M}_{\mathrm{G}}( \mathsf{C})$ {relies} on the underlying graph $\mathrm{G}$ and the set of Boolean interaction rules $\mathsf{C}$. Particularly, we would like to see when $\mathcal{M}_{\mathrm{G}}( \mathsf{C})$ defines an absorbing chain.

{Recall that absorbing states} are the states that can never be left once visited. Therefore, absorbing Markov chains behave fundamentally different with non-absorbing chains.  We introduce two subsets of Boolean mappings:
$$\mathfrak{B}_1=\big\{\mathsf{C}\neq \{\odot_{A}\}\in 2^{\mathsf{H}}:\ \{\odot_{A}\}\subset \mathsf{C} \subseteq \{\odot_2, \odot_3, \odot_{A}, \odot_{B}\}\big\}
$$ and
 $$
 \mathfrak{B}_2=\big\{\mathsf{C}\in 2^{\mathsf{H}}:\ \{\odot_2, \odot_{B}\}\subseteq \mathsf{C}\subseteq \{\odot_2, \odot_3, \odot_{A}, \odot_{B}\}\big\}.
 $$
 We further let $\mathfrak{B}:=\mathfrak{B}_1\mcup\mathfrak{B}_2$.

 Note that there are a total of nine elements in $\mathfrak{B}$. As we show below, Boolean interaction rules in the set   $\mathfrak{B}$ lead to drastically different influences to the absorbing property of the induced chain, compared to the rules outside the set  $\mathfrak{B}$}.

\subsection{Main Results}

We first establish a theorem revealing the connection between the induced Markov chains of {any}  two different   underlying graphs when connectivity is assumed.

\medskip

\begin{theorem}\label{thm:irre}
Suppose $\mathsf{C}\in 2^{\mathsf{H}}\setminus \mathfrak{B}$. {
Then, for any two connected graphs $\mathrm{G}_1$ and $\mathrm{G}_2$ over the node set $\mathrm{V}$, } $\mathcal{M}_{\mathrm{G}_1}( \mathsf{C})$ is an absorbing Markov chain if and only if $\mathcal{M}_{\mathrm{G}_2}( \mathsf{C})$ is an absorbing Markov chain.
\end{theorem}

\medskip

In view of Theorem \ref{thm:irre} and the fact that $\mathrm{G}$ is a connected graph by our standing assumption, whether $\mathcal{M}_{\mathrm{G}}( \mathsf{C})$ being an absorbing chain is fully determined by the interaction rule set $\mathsf{C}$ when $\mathsf{C}$ does not belong to $\mathfrak{B}$.  {Next, we present  the following theorem establishing a necessary and sufficient condition for the induced chain to be absorbing when the Boolean interaction rules come outside the set $\mathfrak{B}$. }

\medskip

\begin{theorem}\label{thm:absorbsn}
Suppose $\mathsf{C}\in 2^{\mathsf{H}}\setminus \mathfrak{B}$. Then $\mathcal{M}_{\mathrm{G}}( \mathsf{C})$ is an absorbing Markov chain if and only if one of the following two  conditions holds
\begin{itemize}
\item[(i)] $\mathsf{C}\subseteq\{\odot_0, \odot_1, \odot_2, \odot_3, \odot_4, \odot_5, \odot_6, \odot_7\}$;
\item[(ii)] $\mathsf{C}\subseteq\{\odot_1, \odot_3, \odot_5, \odot_7, \odot_9, \odot_{B}, \odot_{D}, \odot_{F}\}$.
\end{itemize}
\end{theorem}

\medskip

When the interaction rules  $\mathsf{C}$ indeed comes from the set  $\mathfrak{B}$, the following theorem further gives a tight condition on the absorbing property of the induced chain. Specifically, if $\mathsf{C}$ is one of the nine function sets in $\mathfrak{B}$, the topology of  $\mathrm{G}$ fully determines whether the induced chain is absorbing.

\medskip

\begin{theorem}\label{thm:condition2}
Suppose $\mathsf{C}\in \mathfrak{B}$. Then $\mathcal{M}_\mathrm{G}(\mathsf{C})$ is an absorbing Markov chain if and only if $\mathrm{G}$ does not contain an odd cycle.
\end{theorem}

\medskip

Note that,  Theorem \ref{thm:irre} can actually be inferred from  Theorem \ref{thm:absorbsn}.  Theorem \ref{thm:absorbsn} and Theorem \ref{thm:condition2} together present a comprehensive understanding of the absorbing property of the network Boolean evolution. Below we present two examples illustrating the usefulness of {Theorems~\ref{thm:absorbsn} and~\ref{thm:condition2}}.

{

\medskip

{\em \noindent Example 5.} Consider again the graph $\mathrm{G}$ in Figure \ref{fig:graph2}. With the set of Boolean interaction rules being  $\mathsf{C}=\{\odot_2, \odot_3\}$, the  transition map of the induced Markov chain is illustrated  in Figure~\ref{fig:absorbing-example-thm3}. The chain is absorbing with seven absorbing states: $[0000]$, $[1010]$, $[1001]$ $[1000]$, $[0100]$, $[0010]$, and $[0001]$. This example is  consistent with Theorem \ref{thm:absorbsn}.(i).

\begin{figure}[h]
\centering
 \includegraphics[width=9.8cm]{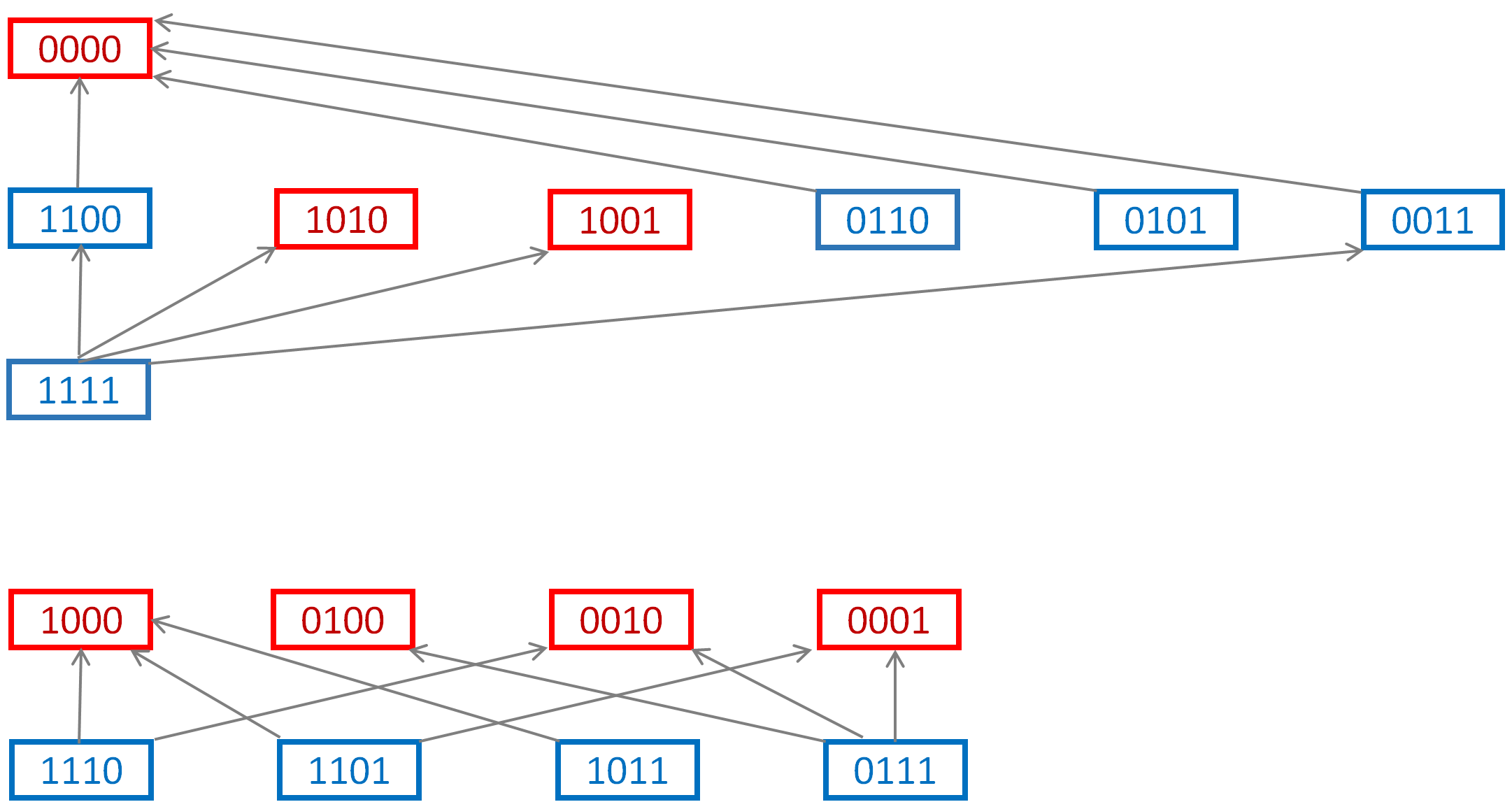}
 \caption{State transitions of the induced Markov chain with $\mathsf{C}=\{\odot_2, \odot_3\}$ for the underlying graph  in Figure \ref{fig:graph2}. The chain is absorbing with seven  absorbing states, which are displayed in red.} \label{fig:absorbing-example-thm3}
 \end{figure}

 {\em \noindent Example 6.} Let the underlying graph $\mathrm{G}$ be given in Figure \ref{fig:graph1}. Let the set of Boolean interaction rules be   $\mathsf{C}=\{\odot_2,\odot_B\}$. The chain is absorbing as shown in Figure \ref{Fig:absorbing-thm4} with two absorbing states $[1001]$ and $[0110]$. This example is consistent with  Theorem \ref{thm:condition2} as the graph does not contain an odd cyle.

\begin{figure}[h]
\centering
 \includegraphics[width=9.5cm]{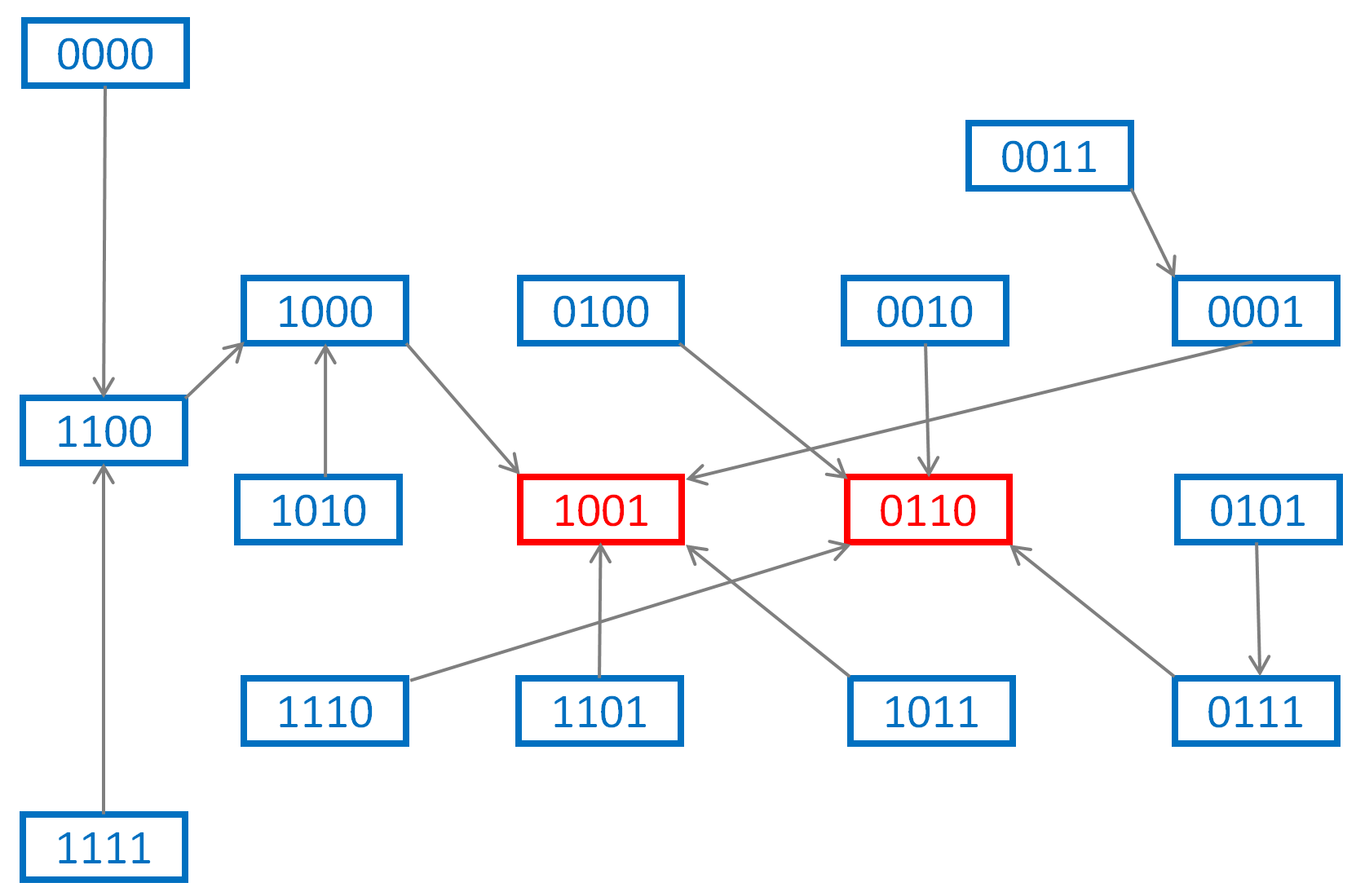}
 \caption{Part of the state transitions of the induced Markov chain with $\mathsf{C}=\{\odot_2,\odot_B\}$ for the four-node cycle  graph  in Figure \ref{fig:graph1}. There exist no outgoing transitions from $[1001]$ and $[0110]$,  revealing that they are absorbing states.   }\label{Fig:absorbing-thm4}
 \end{figure}

\medskip

 {\em \noindent Example 7.} Let the underlying graph $\mathrm{G}$ be given in Figure~\ref{fig:graph2}, and let the set of Boolean interaction rules continue to be   $\mathsf{C}=\{\odot_2,\odot_B\}$. The chain is not absorbing as shown in Figure \ref{Fig:non-absorbing-thm4}, further confirming the conclusion drawn in  Theorem \ref{thm:condition2} as the graph contains an odd cyle.

\begin{figure}[h]
\centering
 \includegraphics[width=9.5cm]{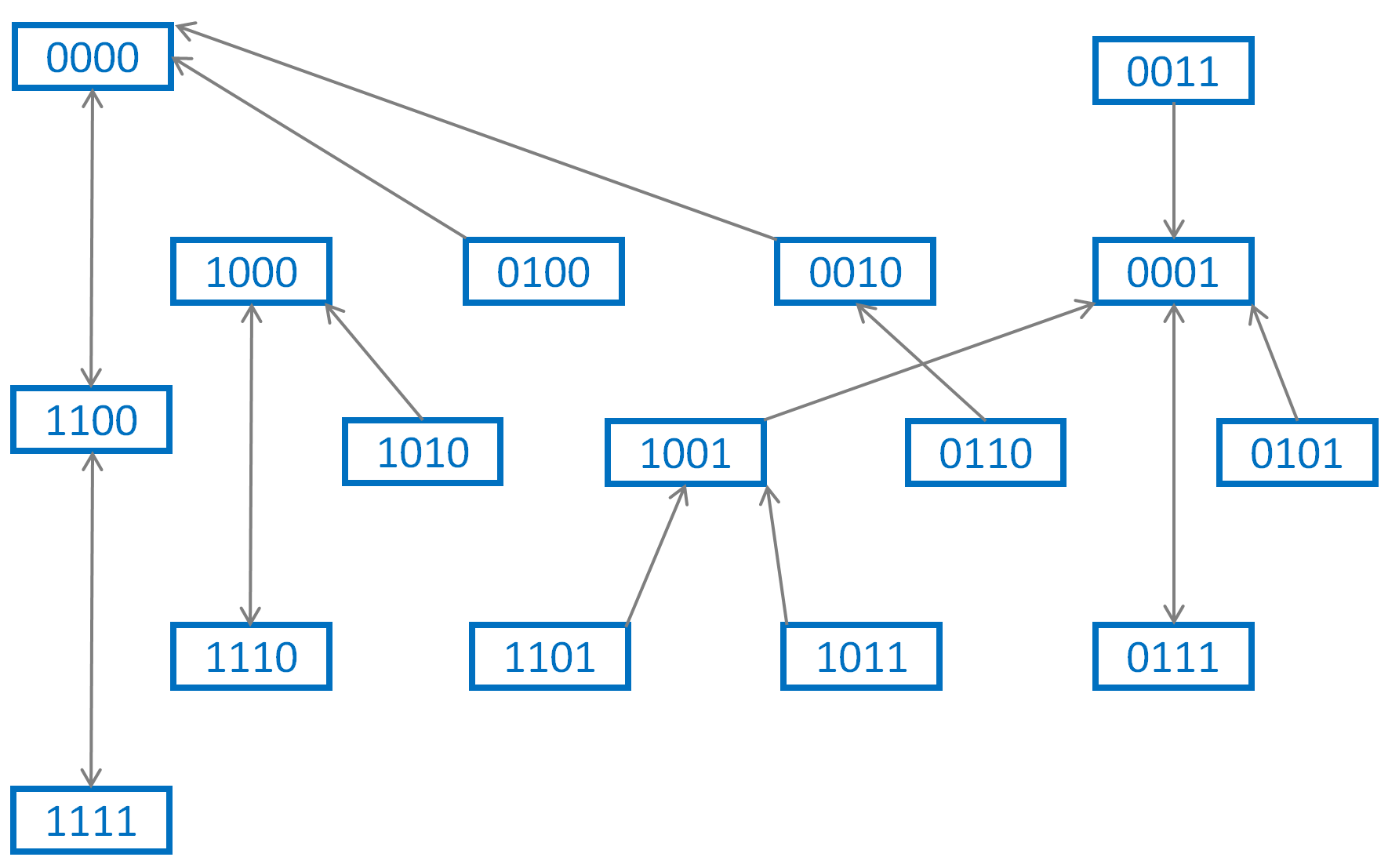}
 \caption{Part of the state transitions of the induced Markov chain with $\mathsf{C}=\{\odot_2,\odot_B\}$ for the underlying graph  in Figure \ref{fig:graph2}, which already shows that the chain cannot be absorbing.   }\label{Fig:non-absorbing-thm4}
 \end{figure}

 }

\subsection{Key Lemma}
To simplify the discussion, we introduce some new notations. For any $S=[s_1 \dots s_n]\in\mathbf{S}_n$, we denote $[S0]$ as $[s_1 \dots s_n0]\in\mathbf{S}_{n+1}$ and $[S1]$ as $[s_1 \dots s_n1]\in\mathbf{S}_{n+1}$. For any $a\in \{0, 1\}$, denote $\overline{a} = 1 - a$. We categorize the states into the following five classes:

$\mathbf{C}_1(\mathrm{G})=\{[s_1 \dots s_{n}]: s_i=0,\ 1\leq i \leq n\}$,

$\mathbf{C}_2(\mathrm{G})=\{[s_1 \dots s_{n}]: s_i=1,\ 1\leq i \leq n\}$,

$\mathbf{C}_3(\mathrm{G})=\{[s_1 \dots s_{n}]: s_i\neq s_j\ \text{for any edge } \{i, j\}\ \text{of }\mathrm{G}\}$,

$\mathbf{C}_4(\mathrm{G})=\{[s_1 \dots s_{n}]$: $\exists i,\ j,\ k$, s.t. $\{i, j\}$ is an edge of $\mathrm{G}$  and  $0=s_i=s_j\neq s_k\}$, and

$\mathbf{C}_5(\mathrm{G})=\{[s_1 \dots s_{n}]$: $\exists i,\ j,\ k$, s.t. $\{i, j\}$ is an edge of $\mathrm{G}$  and  $1=s_i=s_j\neq s_k\}$.

 We may simply write $\mathbf{C}_i$ instead of $\mathbf{C}_i(\mathrm{G})$ whenever this simplification causes no confusion.

In this subsection, we establish a key technical lemma regarding whether a state in the $\mathbf{C}_i$ can be an absorbing state in terms of the selection of $\mathsf{C}$.

\medskip

{
\begin{lemma}\label{lemma-absorbing-state}
(i) The state in $\mathbf{C}_1$ is an absorbing state if and only if
$\mathsf{C} \subseteq \{\odot_0, \odot_1, \odot_2, \odot_3, \odot_4, \odot_5, \odot_6, \odot_7\}$.

(ii) The state in $\mathbf{C}_2$ is an absorbing state if and only if
$\mathsf{C}  \subseteq \{\odot_1, \odot_3, \odot_5, \odot_7, \odot_9, \odot_{B}, \odot_{D}, \odot_{F}\}$.

(iii) A state in $\mathbf{C}_3$  is an absorbing state if and only if
$\mathsf{C}  \subseteq \{\odot_2, \odot_3, \odot_{A}, \odot_{B}\}$.

(iv) A state in $\mathbf{C}_4\setminus \mathbf{C}_5$  is an absorbing state if and only if
$\mathsf{C}  \subseteq \{\odot_2, \odot_3\}$.

(v) A state in $\mathbf{C}_5\setminus \mathbf{C}_4$  is an absorbing state if and only if
$\mathsf{C} \subseteq \{\odot_3, \odot_{B}\}$.

(vi) A state in $\mathbf{C}_5 \mcap\mathbf{C}_4$  is an absorbing state if and only if
$\mathsf{C} \subseteq \{\odot_3\}$.
\end{lemma}
}
{\noindent \it Proof.} (i)  Note that $[0\dots 0]\in \mathbf{C}_1 $ is a state at which any two nodes associated with a common edge must hold the same value $0$. According to the algorithm (\ref{probabilistic-pair}), $[0\dots 0]$ is an absorbing state if and only if for any $\odot_i\in \mathsf{C}$ there holds
$
0 \odot_i 0 = 0.
$
Thus, $[0\dots 0]$ is an absorbing state if and only if
$\mathsf{C}\subseteq \{\odot_0, \odot_1, \odot_2, \odot_3, \odot_4, \odot_5, \odot_6, \odot_7\}$.

{\noindent}(ii) The proof is similar to that in (i), whose details are omitted.

{\noindent}(iii) Let $S\in\mathbf{C}_3$, at which  two nodes sharing a link must hold different values. According to the structure of the algorithmp (\ref{probabilistic-pair}), $S$ is an absorbing state if and only if for any $\odot_i \in \mathsf{C}$, $0 \odot_i 1 = 0$  and $ 1 \odot_i 0 = 1.$
That is, $S$ is an absorbing state, if and only if $\mathsf{C} \subseteq \{\odot_2, \odot_3, \odot_{A}, \odot_{B}\}$.

{
{\noindent}(iv) It is clear that $S\in \mathbf{C}_4\setminus \mathbf{C}_5$ is an absorbing state if and only if for any $\odot_i \in \mathsf{C}$, there hold
\[
0 \odot_i 0 = 0,\ 0 \odot_i 1 = 0 ,\text{ and } 1 \odot_i 0 = 1.
\]
In other words, $S$ is an absorbing state if and only if $\mathsf{C}  \subseteq \{\odot_2, \odot_3\}$.

{\noindent} The proofs of the statements (v) and (vi) are similar to that of (iv), which are, again omitted.  } \hfill$\square$

\subsection{Proof of Theorem \ref{thm:absorbsn}}
This subsection focuses on the proof of Theorem \ref{thm:absorbsn}.

{\noindent} (Necessity.) Assume $\mathcal{M}_\mathrm{G}(\mathsf{C})$ is an absorbing Markov chain.

{
If both $[0\dots 0]\in\mathbf{C}_1$ and   $[1\dots 1]\in\mathbf{C}_2$ are not absorbing,  any state in $\mathbf{C}_4$ or $\mathbf{C}_5$ cannot be absorbing as well according to Lemma \ref{lemma-absorbing-state}(i)-(ii)(iv)-(vi). This leaves the only possibility be that at least one of the states in $\mathbf{C}_3$ is absorbing. Thus, $\mathsf{C}  \subseteq \{\odot_2, \odot_3, \odot_{A}, \odot_{B}\}$ from Lemma \ref{lemma-absorbing-state}(iii).

Next, we conclude that $\mathsf{C}$ can only be  $\{\odot_A\}$ by Lemma \ref{lemma-absorbing-state}(i)-(ii) since $\mathsf{C}\in 2^{\mathsf{H}}\setminus \mathfrak{B}$. However, when $\mathsf{C}=\{\odot_{A}\}$, any state  in $\mathbf{C}_3$  cannot be accessed by any other states. This contradicts the assumption that $\mathcal{M}_\mathrm{G}(\mathsf{C})$ is an absorbing   chain. Therefore, we can only conclude that either $[0\dots 0]\in\mathbf{C}_1$ or  $[1\dots 1]\in\mathbf{C}_2$ is absorbing.

If the state $[0\dots 0]$ is absorbing, we obtain $$
\mathsf{C}\subseteq\{\odot_0, \odot_1, \odot_2, \odot_3, \odot_4, \odot_5, \odot_6, \odot_7\}
$$ according to Lemma \ref{lemma-absorbing-state}(i). While if   $[1\dots 1]$ is absorbing, we have $$
\mathsf{C}  \subseteq \{\odot_1, \odot_3, \odot_5, \odot_7, \odot_9, \odot_{B}, \odot_{D}, \odot_{F}\}
$$ from Lemma \ref{lemma-absorbing-state}(ii). This proves the  necessity statement.

{\noindent} (Sufficiency.) We investigate a few cases.
\begin{itemize}
\item Let $\mathsf{C}\subseteq\{\odot_0, \odot_1, \odot_2, \odot_3, \odot_4, \odot_5, \odot_6, \odot_7\}$. Then $[0\dots 0]$ is absorbing by Lemma \ref{lemma-absorbing-state}(i). We divide the case into a few subcases:
     \begin{enumerate}
        \item If $\mathsf{C} \mcap \{ \odot_0,  \odot_1,  \odot_4,  \odot_5\}\neq \emptyset$, there is a positive probability that $\odot_i\in\{\odot_0,\odot_1,\odot_4,\odot_5\}$ is chosen. Because $1\odot_i 0 = 0,$ any state other than $[1\dots 1]$ can transit to state $[0\dots 0]$ in some finite steps with a positive probability. No matter whether $[1\dots 1]$ is absorbing or not,  $\mathcal{M}_\mathrm{G}(\mathsf{C})$ is an absorbing Markov chain.
        \item Let $\mathsf{C} \mcap \{ \odot_6\}\neq \emptyset$ and consider the update where $\odot_6$ is always selected. Then  for any state  in $\mathbf{C}_4$ or $\mathbf{C}_3$,   two nodes with values $0$ and $1$ respectively  will both hold value $1$ after the interaction, i.e., the network state enters $\mathbf{C}_5$ or $\mathbf{C}_2$. Furthermore, for any state in $\mathbf{C}_5$ or $\mathbf{C}_2$,  two nodes  both holding  value $1$  will both hold  $0$ after the interaction. Thus, for all states in   $\mathbf{C}_2,\dots,\mathbf{C}_5$,  the number of nodes holding value $1$ will be strictly decreasing if $\odot_6$ is always present, until the state transits to  $[0\dots0]$. The chain $\mathcal{M}_\mathrm{G}(\mathsf{C})$ is an absorbing Markov chain since we already know  $[0\dots0]$ is an absorbing state.
        \item Assume $\mathsf{C} = \{\odot_2\}$ or $\mathsf{C} = \{\odot_2,\odot_3\}$ and let $\odot_2$ be  chosen. Then any state in $\mathbf{C}_2$ or $\mathbf{C}_5$ will transit to state in $\mathbf{C}_4\setminus \mathbf{C}_5$ or $\mathbf{C}_1$ in some finite steps.  Thus, $\mathcal{M}_\mathrm{G}(\mathsf{C})$ is an absorbing Markov chain because all states in $\mathbf{C}_1$, $\mathbf{C}_3$, $\mathbf{C}_4\setminus \mathbf{C}_5$ are absorbing  by Lemma \ref{lemma-absorbing-state}.
        \item If $\mathsf{C} = \{\odot_7\}$ or $\mathsf{C} = \{\odot_3,\odot_7\}$, we can use similar discussion in 1) to conclude that any state other than $[0\dots 0]$ can transit to state $[1\dots 1]$ in finite steps. The chain $\mathcal{M}_\mathrm{G}(\mathsf{C})$ is an absorbing Markov chain.
        \item Let $\mathsf{C} = \{\odot_2, \odot_7\}$ or $\mathsf{C} = \{\odot_2,\odot_3,\odot_7\}$. The scenario is similar to 2), where any state can transit to state $[0\dots 0]$ in finite steps.
        \item If $\mathsf{C} = \{\odot_3\}$, all  states are absorbing. Of course $\mathcal{M}_\mathrm{G}(\mathsf{C})$ is an absorbing Markov chain.
     \end{enumerate}
\item Assume $\mathsf{C}\subseteq\{\odot_1, \odot_3, \odot_5, \odot_7, \odot_9, \odot_{B}, \odot_{D}, \odot_{F}\}$. The proof is similar to the case above, whose details are omitted.
\end{itemize}
The proof of Theorem \ref{thm:absorbsn} is now complete.
}

\subsection{Proof of Theorem \ref{thm:condition2}}

In this subsection, we prove Theorem \ref{thm:condition2}.

If $\mathrm{G}$ contains an odd cycle, $\mathbf{C}_3$ is empty. By Lemma \ref{lemma-absorbing-state}, no state in $\mathbf{C}_1\mcup \mathbf{C}_2 \mcup  \mathbf{C}_4 \mcup \mathbf{C}_5$ is absorbing. As $\mathbf{S}_n=\mathbf{C}_1\mcup \mathbf{C}_2 \mcup \mathbf{C}_3 \mcup \mathbf{C}_4 \mcup \mathbf{C}_5 $, no state is absorbing. Thus, $\mathcal{M}_\mathrm{G}(\mathsf{C})$ is not absorbing.
On the other hand, if $G$ does not contain an odd cycle, there are two elements in $\mathbf{C}_3$, and we proceed to prove by induction on the number $n$ of nodes that $\mathcal{M}_\mathrm{G}(\mathsf{C})$ is an absorbing Markov chain.

{For $n=2$}, the conclusion holds straightforwardly. Assume that $\mathcal{M}_\mathrm{G}(\mathsf{C})$ is absorbing for $n=l$. There must be a spanning tree, denoted $\mathrm{G}_{\mathrm{T}_1}$, of $\mathrm{G}$. We further find a subtree $\mathrm{G}_{\mathrm{T}_2}$ of $\mathrm{G}_{\mathrm{T}_1}$ with $\mathrm{G}_{\mathrm{T}_2}$ containing $l$ nodes of $\mathrm{G}_{\mathrm{T}_1}$. Without loss of generality, let $\mathrm{G}_{\mathrm{T}_2}$ contain nodes $1,\dots, l$ of $\mathrm{G}$. By our induction assumption, $\mathcal{M}_{\mathrm{G}_{\mathrm{T}_2}}( \mathsf{C})$ is absorbing.

Now any state in $\mathbf{S}_{l+1}$ can be represented as $[Su]$, where $S\in \mathbf{S}_{l}$ and $u\in\{0,1\}$. As $\mathcal{M}_{\mathrm{G}_{\mathrm{T}_2}}( \mathsf{C})$ is absorbing, there is a positive probability that in finite steps $S$ transits to a state $S^*$ in $\mathbf{C}_3(\mathrm{G}_{\mathrm{T}_2})$. Because $\mathrm{G}_{\mathrm{T}_2}$ is a subgraph of $\mathrm{G}_{\mathrm{T}_1}$, $[Su]$ can transit to $[S^*u]$ in finite steps in $\mathcal{M}_{\mathrm{G}_{\mathrm{T}_1}}( \mathsf{C})$. There will be two cases.

\begin{itemize}
\item If $[S^*u]\in \mathbf{C}_3(\mathrm{G}_{\mathrm{T}_1})$, for $\mathrm{G}$ contains no odd cycle, $[S^*u]\in \mathbf{C}_3(\mathrm{G})$. The proof is done.
\item If $[S^*u]\notin \mathbf{C}_3(\mathrm{G}_{\mathrm{T}_1})$, there must be some node $j$ associated with node $l+1$ over graph $\mathrm{G}_{\mathrm{T}_1}$. Because $\mathsf{C}\in \mathfrak{B}$, there is a positive probability that $\odot_{A}$ or $\odot_{B}$ is chosen. Note that
$
0 \odot_{A} 0 = 1$, $0 \odot_{B} 0 = 1
$, $1\odot_{A} 1 = 0$ and $1 \odot_{2} 1 = 0$.
Thus, by (\ref{probabilistic-pair}), $[S^*u]$ transits to $[S^*\overline{u}]$ with positive probability in $\mathcal{M}_{\mathrm{G}_{\mathrm{T}_1}}(\mathsf{C})$. Moreover, $[S^*\overline{u}]\in \mathbf{C}_3(\mathrm{G}_{\mathrm{T}_1})$.  For $\mathrm{G}$ contains no odd cycle, $[S^*\overline{u}]\in \mathbf{C}_3(\mathrm{G})$ leads to the desired result.
\end{itemize}

The proof of Theorem \ref{thm:condition2} is completed.

\section{Conclusions}\label{sec:conclusion}

We proposed and investigated a Boolean gossip model, which may be useful in describing social opinion evolution as well as serves as a simplified probabilistic Boolean network.  With positive node interactions, it was shown that the node states asymptotically converge to a consensus  represented by a binary random variable, whose  distribution {was} studied  for large-scale complete networks in light of mean-field approximation methods. By combinatorial analysis the  number of communication classes of the positive Boolean network was counted against the topology of the underlying interaction graph. With general Boolean interaction rules, the emergence of  absorbing network Boolean dynamics was explicitly characterized by the network structure. It turned out that local structures in terms of existence of cycles can drastically change fundamental  properties of the Boolean network. In future, it will be interesting to look into the possibility of extending the graphical analysis established in the current work to multi-state  Boolean networks \cite{m1,m2} where each node may hold a state from a finite set with more than two values.

\medskip

\medskip

\section*{Appendix. Proof of Theorem \ref{thm:class}}
For each $n$, we use ${\rm Mod}_n(i)$ to denote the unique integer $j$ satisfying $1\leq j\leq n$ and $i\equiv j\mod(n)$. Recall that for any $a\in \{0, 1\}$, we denote $\overline{a} = 1 - a$.

We  prove the statements of  Theorem \ref{thm:class} in a few steps starting with a few fundamental graphs.

\subsection*{A.1 Line graph}

In this subsection we prove Theorem \ref{thm:class}.(i) stating that $\chi_{_{\mathsf{C}_{\rm pst}}}(\mathrm{G})=2n$ when $\mathrm{G}$ is  a line  graph. Without loss of generality we assume the edges of  $\mathrm{G}$ are $ \{i,i+1\}$ for $i=1,\dots,n-1$. The proof is outlined as follows. We first introduce the notion of {\it $\mathpzc{L}$-reduced state} for each state in $\mathbf{S}_n$. Then, we prove that any two states communicate with each other  if and only if their $\mathpzc{L}$-reduced states are identical. Finally, we count the number of  $\mathpzc{L}$-reduced states in the state space and therefore obtain the number of communication classes.

\medskip

\begin{definition}($\mathpzc{L}$-reduced states) Let $[s_1 \dots s_n]\in\mathbf{S}_n$. There exists a unique  partition of $s_1, \dots , s_n$ into
\[
\begin{array}{ll}
s_1=\dots=s_{i_1}=r_1,& i_1\geq 1;\\
s_{i_1+1}=\dots=s_{i_2}=r_2,& i_2> i_1;\\
 \dots  & \\
s_{i_{d-2}+1}=\dots=s_{i_{d-1}}=r_{d-1},& i_{d-1}> i_{d-2};\\
s_{i_{d-1}+1}=\dots=s_{n}=r_d
\end{array}
\]
such that $r_i\neq r_{i+1}$ for all $i=1, \dots , d-1$. Then $[r_1 \dots  r_d]:=\mathpzc{L}([s_1 \dots  s_n])$ is termed the $\mathpzc{L}$-reduced states of $[s_1 \dots  s_n]$.
\end{definition}

\medskip

Note that the values of any two consecutive elements in an $\mathpzc{L}$-reduced state are different.
The following two lemmas hold.

\medskip

\begin{lemma}\label{lemma-assess} Suppose $\mathrm{G}$ is  a line  graph.
Then $\mathpzc{L}([s_1 \dots  s_n])$ is a subsequence of $\mathpzc{L}([q_1 \dots  q_n])$ if  $[s_1 \dots s_n]$ is accessible from $[q_1 \dots q_n]$. More precisely, denoting  $$
\mathpzc{L}([s_1 \dots  s_n])= [r_1 \dots  r_d], \
\mathpzc{L}([q_1 \dots  q_n])=[h_1 \dots  h_{d'}]
 $$
  there holds $d\leq d'$, and moreover,  there exist $1\leq\tau_1<\tau_2< \dots <\tau_d\leq d'$ such that $r_{i}=h_{\tau_i}$ for all $i=1,\dots, d$.
\end{lemma}
{\it Proof.} By the definition of accessibility, there is a nonnegative integer $t$ such that
\[
\mathbb{P}\big(X_t=[s_1 \dots  s_n]\ \big|\ X_0=[q_1 \dots  q_n]\big)>0.
\]

First we assume  $t=1$. According to  the structure of (\ref{probabilistic-pair}),  either $[s_1 \dots s_n]=[q_1 \dots q_n]$, or there is  $u\in\{1, \dots , n\}$ such that  $s_u\neq q_u$ and $s_i=q_i$ for all $i\neq u$. The desired conclusion  obviously holds   if $[s_1 \dots s_n]=[q_1 \dots q_n]$. For the latter case, there is $q_{v}$ with $v=u+1$ or $v=u-1$ such that
$
q_u\neq q_v.
$ Consequently, the two states  $[s_1 \dots s_n]$ and $[q_1 \dots q_n]$ differ with each other only at $s_u$ and $q_u$ and satisfy
$$
 s_u \neq q_u,\  s_u=s_v, \ q_u\neq q_v.
$$
Then it is easy to verify that  $\mathpzc{L}([s_1 \dots  s_n])$ is a subsequence of $\mathpzc{L}([q_1 \dots  q_n])$ from the definition of $\mathpzc{L}$-reduced states.

Now we proceed to let $t=2$. There will be a state $[w_1 \dots  w_n]$ such that $[s_1 \dots s_n]$  is one step accessible from $[w_1 \dots  w_n]$, and $[w_1 \dots  w_n]$ is one step accessible from $[q_1 \dots  q_n]$. Utilizing the above understanding for the case with $t=1$ we know  $\mathpzc{L}([s_1 \dots  s_n])$ is a subsequence of $\mathpzc{L}([w_1 \dots  w_n])$ and
 $\mathpzc{L}([w_1 \dots  w_n])$ is a subsequence of $\mathpzc{L}([q_1 \dots  q_n])$, which in turn imply    $\mathpzc{L}([s_1 \dots  s_n])$ is a subsequence of $\mathpzc{L}([q_1 \dots  q_n])$. Therefore the desired conclusion holds for $t=2$. Apparently the argument can be recursively carried out and the result holds for arbitrary integer  $t$. We have now completed the proof of the lemma.  \hfill$\square$

\medskip

\begin{lemma}\label{lemma-line}Let $\mathrm{G}$ be a line graph and consider $S=[s_1 \dots s_n], Q=[q_1\dots q_n]\in \mathbf{S}_n$. Then $S$ and $Q$ communicate with each other if and only if they have identical $\mathpzc{L}$-reduced states.
\end{lemma}
{\it Proof.}  The necessity part of this lemma follows directly from Lemma \ref{lemma-assess}.  In the following we focus only on the sufficiency part. Let the identical $\mathpzc{L}$-reduced state of $[s_1 \dots s_n]$ and $[q_1 \dots q_n]$ be $[r_1\dots r_l]$. We carry out an induction argument on $l$ for any $n\geq l$.

Let $l=1$. Then  $[0\dots0]_n$ and $[1 \dots 1]_n$   are the  two possible  states   for  $[s_1 \dots s_n]$ and $[q_1 \dots q_n]$.  The desired conclusion holds straightforwardly. Now assume:

\medskip

 \noindent {\it Induction Hypothesis}: The statement of the lemma holds true for all $l\leq k$ and all $n\geq l$.

\medskip

We proceed to prove the statement for $l=k+1$ and $n\geq l$. Denote $i_1=\max\{h:r_1=s_i,\ 1\leq i\leq h\}$ and $j_1=\max\{h:r_1=q_i,\ 1\leq i\leq h\}$.
By symmetry we may assume $i_1\leq j_1$ and we use the following two observations:
\begin{itemize}
\item[{\it a)}]  The state $[q_1 \dots q_n]$ communicates with the state $$
[q_1 \dots  q_{i_1}\overline{q}_{i_1+1} \dots  \overline{q}_{j_1} q_{j_1+1} \dots  q_n]
 $$
 by the definition of $j_1$.
 \item[{\it b)}]
 The two states $[\overline{q}_{i_1+1} \dots  \overline{q}_{j_1} q_{j_1+1} \dots  q_n]$ and $[s_{i_1+1}s_{i_1+2} \dots  s_n]$ have the same $\mathpzc{L}$-reduced state $[r_2 \dots  r_{l}]$. Therefore by our induction hypothesis, $[\overline{q}_{i_1+1} \dots  \overline{q}_{j_1} q_{j_1+1} \dots  q_n]$ and $[s_{i_1+1} \dots  s_n]$ communicate with each other, which in turn yields that $[s_{1} \dots  s_n]$ communicates with $$
    [q_1 \dots  q_{i_1}\overline{q}_{i_1+1} \dots  \overline{q}_{j_1} q_{j_1+1} \dots  q_n].
    $$
\end{itemize}
Combining $a)$ and $b)$ we immediately know that    $[s_{1} \dots  s_n]$ communicates  with  $[q_1 \dots q_n]$. By the principle of mathematical induction we have completed the proof of the lemma. \hfill$\square$

We are now ready to count the number of communication classes for the line graph, which equals to  the number of $\mathpzc{L}$-reduced states according to Lemma \ref{lemma-line}. For each $m=1,\dots,n$, there are two different $\mathpzc{L}$-reduced states with length $m$, i.e., $[r_1  \dots  r_{m}]$ with $r_1=0$ or $r_1=1$.
Consequently,  there are a total of  $2n$ different $\mathpzc{L}$-reduced states. This concludes the proof for Theorem \ref{thm:class}.(i).

\subsection*{A.2 Cycle  graph}

In this subsection, we prove  the case with $\mathrm{G}$ being a cycle graph. Without loss of generality, let $\mathrm{G}$ be the cycle graph with edges $\big\{i,{\rm Mod}_n(i+1)\big\}$, $i=1,\dots,n$.

We introduce  some useful notations that will be used subsequently. For any $k$, we use $\sigma_k$ to denote the permutation on set $\{1,\dots,k\}$ with $\sigma_k(i)={\rm Mod}_k(i+1)$ for $i=1,\dots,k$. We further define  $\mathpzc{P}_{\sigma_k}$ as a mapping over $\mathbf{S}_{k}$ by
\[\mathpzc{P}_{\sigma_k}([s_1 \dots s_k])=[s_{\sigma_{k}(1)} \dots s_{\sigma_{k}(k)}]\]
for all $[s_1 \dots s_k]\in \mathbf{S}_{k}$.  Intuitively, if we place these $k$ nodes uniformly on a cycle and denote the value of each node on them, then the result of $\mathpzc{P}_{\sigma_k}$ on a state is obtained by rotating all the values counterclockwise. We also define a mapping ${f}_{[k_1,k_2]}$ over $\mathbf{S}_{n}$ by that for any $[t_1\dots t_n]$, ${f}_{[k_1,k_2]}\big([t_1\dots t_n] \big)=[r_1\dots r_n]$ with $r_i=t_i$ for $i\neq k_2$ and $r_i=t_{k_1}$ for $i = k_2$.

\medskip

\begin{definition}($\mathpzc{K}$-reduced  states) Let $[s_1 \dots s_n]\in\mathbf{S}_n$ with $[r_1 \dots  r_d] = \mathpzc{L}([s_1 \dots  s_n])$ being its $\mathpzc{L}$-reduced states. The $\mathpzc{K}$-reduced  states of $[s_1 \dots s_n]\in\mathbf{S}_n$, denoted $\mathpzc{K}([s_1 \dots  s_n])$, is defined as follows:{
\[ \mathpzc{K}([s_1 \dots  s_n]) =
  \begin{cases}
    [r_1]     & \quad \text{if } d=1;\\
   [r_1 \dots  r_d]  & \quad \text{if $d>1$ and $r_d\neq r_1$};\\
   [r_1 \dots  r_{d-1}] & \quad \text{if $d>1$ and $r_d=r_1$}.\\
  \end{cases}
\]}
\end{definition}
\medskip

Let $|\mathpzc{K}(S)|$ be the number of digits in $\mathpzc{K}(S)$ for $S\in \mathbf{S}_n$.
According to the definition, the values of any two consecutive  elements of $\mathpzc{K}$-reduced  states are different. Moreover, if there are at least two entries  of $\mathpzc{K}$-reduced  states, the first entry  is different from the last one. The following lemma can be established using a similar analysis as we used in Lemma \ref{lemma-assess}.

\medskip

\begin{lemma}\label{cycle-lemma-length}
Suppose $\mathrm{G}$ is a cycle graph,

(i) $|\mathpzc{K}(S)|$ is either $1$ or an even integer;

(ii) If $d$ is one or an even integer, then there is $S\in \mathbf{S}_n$ with $|\mathpzc{K}(S)|=d$.

(iii) If $S$ is accessible from  $T$, then $|\mathpzc{K}(S)|\leq |\mathpzc{K}(T)|$.
\end{lemma}
\medskip

\begin{lemma}\label{cycle-lemma-same-length}
Consider $S, T\in \mathbf{S}_{n}$. If $1<|\mathpzc{K}(S)|=|\mathpzc{K}(T)|<n$, then $S$ and $T$ communicate with each other.
\end{lemma}
{\noindent \it Proof.} Denote $S=[s_1\dots s_n]$ and $T=[t_1\dots t_n]$. We prove this lemma in a few steps.

{\noindent Step 1.} We first prove that $S$ communicates with $\mathpzc{P}_{\sigma_n}^l(S)$ for any integer $l$ if $|\mathpzc{K}(S)|<n$.
Note that if $|\mathpzc{K}(S)|=1$, $S$ must be $[0 \dots 0]_{n}$ or $[1\dots 1]_{n}$. The claim holds straightforwardly.

Now we assume $|\mathpzc{K}(S)|>1$. Since $|\mathpzc{K}(S)|<n$, the set $$\mathcal{I}:=\{i: s_i=s_{{\rm Mod}_n(i+1)},\ 1\leq i\leq n\}$$ is nonempty. Moreover, because $|\mathpzc{K}(S)|>1$, we can find $j\in  \mathcal{I}$ such that $s_{{\rm Mod}_n(j+1)}\neq s_{{\rm Mod}_n(j+2)}$. By the structure of (\ref{probabilistic-pair}), the state ${f}_{[{\rm Mod}_n(j+2), {\rm Mod}_n(j+1)]}(S)$ is accessible from $S$. By the definition of $j$, there holds $${f}_{[j,{\rm Mod}_n(j+1)]}{f}_{[{\rm Mod}_n(j+2),{\rm Mod}_n(j+1)]}(S)=S.$$
That is to say, the state $S$ is accessible from $$
{f}_{[{\rm Mod}_n(j+2), {\rm Mod}_n(j+1)]}(S).
$$
Therefore, $S$ communicates with  ${f}_{[{\rm Mod}_n(j+2), {\rm Mod}_n(j+1)]}(S)$. Applying this argument recursively, we obtain that $S$ communicates with
$$
{f}_{[{\rm Mod}_n(j+n), {\rm Mod}_n(j+n-1)]}\dots{f}_{[{\rm Mod}_n(j+2), {\rm Mod}_n(j+1)]  }(S),
$$
a state  equal to $\mathpzc{P}_{\sigma_n}(S)$. It is then convenient to conclude that $S$ communicates with $\mathpzc{P}_{\sigma_n}^l(S)$ for any integer $l$.

{\noindent Step 2.} In this step, we prove that if $S=[s_1\dots s_n]$ and $T=[t_1\dots t_n]$ have identical $\mathpzc{K}$-reduced states, then $S$ and $T$ communicate with each other. Let $\mathpzc{K}(S)=\mathpzc{K}(T)=[c_1\dots c_d]$.
If $d=1$ or $n$,  it is easy to see $S=T$. Now assume $1<d<n$.

Because $d>1$, the sets $\{i:s_i\neq s_0\}$ and $\{i:t_i\neq t_0\}$ are not empty. Denote $j_1=\max\{i:\ s_i\neq s_0\}$, and $j_2=\max\{i:\ t_i\neq t_0\}$. Without loss of generality we assume $j_1>j_2$. Apparently  $T$ communicates with ${f}_{[j_2, j_2+1]}(T)$. Further we know that $T$ communicates with $$
T^*=[t^*_1 \dots t^*_{n}]:={f}_{[j_1-1, j_1]}\dots {f}_{[j_2+1, j_2+2]}{f}_{[j_2, j_2+1] }(T).
$$ Moreover, we can conclude that $j_1=\max\{i:\ t^*_i\neq t^*_0\}$, and $T^*$ and $T$ have the same $\mathpzc{K}$-reduced state. So $T^*$ and $S$ have the same $\mathpzc{K}$-reduced   state. By the definition of $\mathpzc{K}$-reduced   state and the fact that $j_1=\max\{i:\ s_i\neq s_0\}=\max\{i:\ t^*_i\neq t^*_0\}$, we know that the $\mathpzc{L}$-reduced state of $S$ is equal to the $\mathpzc{L}$-reduced state of $T^*$. Define a new line graph $\tilde{\mathrm{G}}$, whose nodes are the nodes of $\mathrm{G}$ with edges being $\{i, i+1\}$ for $i=1,\dots n-1$. According to Lemma \ref{lemma-line}, $S$ also communicates with $T^*$ in $\mathcal{M}_{\tilde{\mathrm{G}}}(\mathsf{C}_{\rm pst})$. Therefore, $S$ communicates with $T^*$ in $\mathcal{M}_\mathrm{G}(\mathsf{C}_{\rm pst})$, because $\tilde{\mathrm{G}}$ is a subgraph of $\mathrm{G}$. Thus, $S$ and $T$ communicate with each other.

{\noindent Step 3.} This step will complete the proof.

Let $d = |\mathpzc{K}(S)|$.
If $\mathpzc{K}(S)=\mathpzc{K}(T)$, we have known that $S$ and $T$ communicate with each other. We only need to consider the case $\mathpzc{K}(S)\neq \mathpzc{K}(T)$. Because $|\mathpzc{K}(S)|=|\mathpzc{K}(T)|$, there must hold that $\mathpzc{K}(T)=\mathpzc{P}_{\sigma_d}(\mathpzc{K}(S))$. For $d>1$, the set $\{i:s_i\neq s_0\}$ is nonempty. Define $j=\min\{i:s_i\neq s_0\}$. According to Step 1, $S$ communicates with $\mathpzc{P}_{\sigma_n}^{j-1}(S)$. By the definition of $\mathpzc{K}$-reduced states, we know that the $\mathpzc{K}$-reduced  state of $\mathpzc{P}_{\sigma_n}^{j-1}(S)$ is $\mathpzc{P}_{\sigma_d}(\mathpzc{K}(S))$, i.e., $\mathpzc{K}(T)$. Therefore, $\mathpzc{P}_{\sigma_n}^{j-1}(S)$ communicates with $T$, implying that $S$ communicates with $T$.\hfill$\square$

\medskip

Now, we are ready to count the number of communication classes.
According to Lemma \ref{cycle-lemma-length}, the digit number $d$ of the $\mathpzc{K}$-reduced  states of all the states in the same  communication class are identical. Moreover, $d$ can be $1$ or even numbers.
If $n=2m$, there are three cases:
\begin{itemize}
\item[(i)] For $d=1$, there are two communication classes $\{[0\dots 0]\}$ and $\{[1\dots 1]\}$.

\item[(ii)] For each $d=2, 4, \dots, 2m-2$, according to Lemma \ref{cycle-lemma-length} and Lemma \ref{cycle-lemma-same-length}, there is a unique  communication class whose elements have $\mathpzc{K}$-reduced with $d$ digits.
\item[(iii)] For $d=2m$, the two states $S_0:=[s_1\dots s_{2m}]$ and $T_0:= [\bar{s}_1\dots \bar{s}_{2m}]$ with $s_{2i-1}=1$ and $s_{2i}=0$ for $i=1,\dots,m$,  are the only states whose $\mathpzc{K}$-reduced  states are of length $2m$. Moreover, either  $S_0$ or $ T_0$ cannot be accessible from any other state. That is to say, they form two communication classes.
\end{itemize}
As a result, there are a total of $m+3$ communication classes. We have completed the proof for the case $n=2m$. The case with $n=2m+1$ can be similarly analyzed, whose detailed proof is omitted. This concludes the proof of Theorem \ref{thm:class}(ii).

\subsection*{A.3 Star graph}

In this subsection, we prove that $\chi_{_{\mathsf{C}_{\rm pst}}}(\mathrm{G})=5$ if $\mathrm{G}$ is a star graph with $n(\geq 4)$ nodes. Note that a connected graph is called a star graph
if there is a node such that all the edges of the graph contain this node. This particular node is called the center node of the graph.

The following proposition characterizes the communication classes for $\mathcal{M}_\mathrm{G}(\mathsf{C}_{\rm pst})$ over a star graph $\mathrm{G}$.

\medskip

\begin{proposition}\label{star}
Let $\mathrm{G}$ be a star graph with $n(\geq 4)$ nodes. Then $\chi_{_{\mathsf{C}_{\rm pst}}}(\mathrm{G})=5$. Moreover, letting node $1$ be the center node, the five classes are

$\mathbf{F}_n^1=\{[s_1 \dots s_{n}]: s_i=0,\ 1\leq i \leq n\}$,

$\mathbf{F}_n^2=\{[s_1 \dots s_{n}]: s_i=1,\ 1\leq i \leq n\}$,

$\mathbf{F}_n^3=\{[s_1 \dots s_{n}]: s_1=0,\ s_i=1,\ 2\leq i \leq n\}$,

$\mathbf{F}_n^4=\{[s_1 \dots s_{n}]: s_1=1,\ s_i=0,\ 2\leq i \leq n\}$,

$\mathbf{F}_n^5=\{[s_1 \dots s_{n}]: \exists i,\ j,\ 2\leq i,\ j\leq n,\ s_i=0,\ s_j=1\}$.

\end{proposition}
{\it Proof.}  Denote $S_n^{\langle 1 \rangle},\dots, S_n^{\langle 4\rangle}$ as the singleton state in $\mathbf{F}_n^1,\dots, \mathbf{F}_n^4$, respectively. Moreover, any other state cannot be accessible from $S_n^{\langle 1 \rangle}$ or $S_n^{\langle 2\rangle}$, while $S_n^{\langle 3\rangle}$ or $S_n^{\langle 4\rangle}$ cannot be accessible from any other state. Thus, they do form communication classes, respectively. We only need to prove all the elements in $\mathbf{F}_n^5$ communicate with each other. We prove this by induction.

First, let $n=4$. There are $12$ elements of $\mathbf{F}_4^5$, listed as $[0100]$, $[0010]$, $[0001]$, $[0011]$, $[0101]$, $[0110]$, $[1100]$, $[1010]$, $[1001]$, $[1011]$, $[1101]$, $[1110]$. It is easy to verify that they are in the same communication class.

Assume that for $n=k\geq 4$, all the elements in $\mathbf{F}_n^5$ communicate with each other. Now we prove the case for $n=k+1$. Let $\mathrm{G}$ be a star graph with $k+1$ nodes with node $1$ being its center node. Let $\mathrm{G}^*$ be the subgraph of $\mathrm{G}$ with nodes $1,\dots, k$ and all edges containing them in $\mathrm{G}$. In fact, $\mathrm{G}^*$ is a star graph with $k$ nodes. By our induction assumption, all elements in $\mathbf{F}_k^5$ communicate with each other in $\mathcal{M}_\mathrm{G}(\mathsf{C}_{\rm pst})$. Because $\mathrm{G}^{*}$ is a subgraph of $\mathrm{G}$, all elements in $\mathbf{A}: = \{[S0]\in \mathbf{S}_{k+1}:S\in \mathbf{F}_k^5\}$ communicate with each other, and all elements in $\mathbf{B}:=\{[S1]\in \mathbf{S}_{k+1}:S\in \mathbf{F}_k^5\}$ communicate with each other.

Note that
$$\mathbf{F}_{k+1}^5=\mathbf{A}\mcup\mathbf{B}\mcup\{[S_k^{\langle 1 \rangle}1],[S_k^{\langle 2\rangle}0],[S_k^{\langle 3\rangle}0],[S_k^{\langle 4\rangle}1]\}.$$
Introduce $U_k^{\langle a\rangle}=[010\dots 0]$, $U_k^{\langle b\rangle}=[101\dots 1]$, $U_k^{\langle c \rangle}=[001\dots1]$ and $U_k^{\langle d \rangle}=[110\dots 0]$. They are elements of $\mathbf{F}_k^5$. It is easy to verify that $[U_k^{\langle a\rangle}0]\in \mathbf{A}$ is accessible from $[U_k^{\langle a\rangle}1]\in \mathbf{B}$. Moreover, $[U_k^{\langle b\rangle}1]\in \mathbf{B}$ is accessible from $[U_k^{\langle b\rangle}0]\in \mathbf{A}$. Therefore, all elements in $\mathbf{A}\mcup\mathbf{B}$ communicate with each other.

It is straightforward to verify that  $[S_k^{\langle 2\rangle}0]$ communicates with $[S_k^{\langle 3\rangle}0]$. Also, $[U_k^{\langle c \rangle}0]\in\mathbf{A}$ is accessible from $[S_k^{\langle 3\rangle}0]$ and $[S_k^{\langle 2\rangle}0]$ is accessible from $[U_k^{\langle b\rangle}0]\in \mathbf{A}$. Thus,
all elements in $\mathbf{A}\mcup \{[S_k^{\langle 2\rangle}0],[S_k^{\langle 3\rangle}0]\}$ communicate with each other.
Moreover, $[S_k^{\langle 1 \rangle}1]$ communicates with $[S_k^{\langle 4\rangle}1]$, $[U_k^{\langle d \rangle}1]\in\mathbf{B}$ is accessible from $[S_k^{\langle 4\rangle}1]$, and $[S_k^{\langle 1 \rangle}1]$ is accessible from $[U_k^{\langle a\rangle}1]\in\mathbf{B}$. Therefore, all elements in $\mathbf{B}\mcup \{[S_k^{\langle 1 \rangle}1],[S_k^{\langle 4\rangle}1]\}$ communicate with each other.
Summarizing all these relations we know all elements in
$$\mathbf{F}_{k+1}^5=\mathbf{A}\mcup\mathbf{B}\mcup\{[S_k^{\langle 1 \rangle}1],[S_k^{\langle 2\rangle}0],[S_k^{\langle 3\rangle}0],[S_k^{\langle 4\rangle}1]\}$$
communicate with each other. This completes the proof of this proposition.\hfill$\square$

\subsection*{A.4 Tree}
The following result presents a characterization of the number of communication classes for tree graph that is not a line.

\medskip

\begin{proposition}\label{prop:tree}
Let $\mathrm{G}$ be a tree, having at least one node with degree greater than $2$, i.e., $\mathrm{G}$ is not a line graph. Then $\chi_{_{\mathsf{C}_{\rm pst}}}(\mathrm{G})=5$. The five communication classes can be described as follows:

$\mathbf{J}_n^1=\big\{[s_1  \dots  s_{n}]:\ s_i=0,\ 1\leq i \leq n\big\}$,

$\mathbf{J}_n^2=\big\{[s_1  \dots  s_{n}]:\ s_i=1,\ 1\leq i \leq n\big\}$,

$\mathbf{J}_n^3=\big\{[s_1  \dots  s_{n}]:$ $s_1=0,\ s_i\neq s_j$ \text{for any edge } $\{i, j\}$ \text{of} $\mathrm{G}\big\}$,

$\mathbf{J}_n^4=\big\{[s_1  \dots s_{n}]:\ s_1=1,\ s_i\neq s_j$ \text{for any edge } $\{i, j\}$ \text{of} $\mathrm{G}\big\}$, and

$\mathbf{J}_n^5=\big\{[s_1  \dots  s_{n}]:$ $\exists i,\ j,\ k$, s.t. $\{i, j\}$ is an edge of $\mathrm{G}$ \text{and} $s_i=s_j\neq s_k\big\}$.
\end{proposition}

\noindent{\it Proof}.
It is straightforward to verify that any of $\mathbf{J}_n^1$, $\mathbf{J}_n^2$, $\mathbf{J}_n^3$, $\mathbf{J}_n^4$ contains a unique element, and forms a communication class. We now prove $\mathbf{J}_n^5$ is a communication class using an induction argument.

For $n=4$, $\mathrm{G}$ is a star graph which is proved in Proposition \ref{star}. Now assume that this proposition holds for $n=l\geq 4$.

For any tree $\mathrm{G}$ with $l+1$ nodes that is not a line graph, there is a subgraph $\mathrm{G}^{*}$ with $l$ nodes which is still a tree. Without loss of generality, we denote the node not in $\mathrm{G}^{*}$ as node $v_*=l+1\in \mathrm{V}$. We use $v_0$ to denote the node with the highest degree in $\mathrm{G}$ (If there are more than one such nodes, we just choose one of them arbitrarily). There is a path $(v_0,\ v_1,\  \dots ,\ v_h,\ v_*)$ connecting node $v_0$ and node $v_*$ in $\mathrm{G}$, where $h\geq 0$ is an integer.

By the induction assumption, the communication classes of $\mathcal{M}_{\mathrm{G^*}}( \mathsf{C})$ are $\mathbf{J}_l^1, \dots, \mathbf{J}_l^5$ with each $\mathbf{J}_l^k$ defined by replacing $n$ with $l$ in $\mathbf{J}_n^k$. Denote $\mathbf{A} = \{[S0]\in \mathbf{S}_{l+1}:S\in \mathbf{J}_l^5\}$ and $\mathbf{B} = \{[S1]\in \mathbf{S}_{l+1}:S\in \mathbf{J}_l^5\}$. Note that
$$\mathbf{J}_{l+1}^5 = \mathbf{A}\mcup\mathbf{B}\mcup\{[S_l^{\langle 1\rangle}1],[S_l^{\langle 2\rangle}0],[S_l^{\langle 3\rangle}0],[S_l^{\langle 4\rangle}1]\}.$$
Because $\mathrm{G}^{*}$ is a subgraph of $\mathrm{G}$, all elements in $\mathbf{A}$ communicate with each other, and all elements in $\mathbf{B}$ communicate with each other. Note that if $\mathrm{G}^{*}$ is a star graph with the $v_h$ being the center node, $\mathrm{G}$ will be a star graph. This falls to the case discussed in Proposition \ref{star}. We assume $\mathrm{G}^{*}$ is not a star graph for the remainder of the proof.

Introduce $$U_l^{\langle a\rangle}=
[0\dots0\underset{\underset{v_h}{\uparrow}}{1}0\dots 0]
,\ \ U_l^{\langle b\rangle}=
[1\dots1\underset{\underset{v_h}{\uparrow}}{0}1\dots 1].
$$ We have $U_l^{\langle a\rangle}$, $U_l^{\langle b\rangle}\in\mathbf{J}_l^5$.
It is easy to verify that $[U_l^{\langle a\rangle}1]\in \mathbf{B}$ is accessible from $[U_l^{\langle a\rangle}0]\in \mathbf{A}$. Moreover, $[U_l^{\langle b\rangle}0]\in \mathbf{A}$ is accessible from $[U_l^{\langle b\rangle}1]\in \mathbf{B}$. Therefore, all elements in $\mathbf{A}\mcup\mathbf{B}$ communicate with each other.

We further denote $S_l^{\langle 4\rangle}=[\gamma_1\dots \gamma_l]$ and $S_l^{\langle 3\rangle}=[\beta_1\dots \beta_l]$, and then $U_l^{\langle c\rangle}:=[\gamma_1\dots \gamma_{v_h-1}\overline{\gamma_{v_h}}\gamma_{v_h+1}\dots \gamma_l]$,\\ $U_l^{\langle d\rangle}:=[\beta_1\dots \beta_{v_h-1}\overline{\beta_{v_h}}\beta_{v_h+1}\dots \beta_l]$.
It is straightforward to verify that $[S_l^{\langle 1\rangle}1]$ communicates with $[U_l^{\langle a\rangle}1]\in\mathbf{B}$, $[S_l^{\langle 2\rangle}0]$ communicates with $[U_l^{\langle b\rangle}0]\in\mathbf{A}$, $[S_l^{\langle 3\rangle}0]$ communicates with $[U_l^{\langle d\rangle}0]\in\mathbf{A}$, and $[S_l^{\langle 4\rangle}1]$ communicates with $[U_l^{\langle c\rangle}1]\in\mathbf{B}$. Thus, any two elements in $\mathbf{J}_{l+1}^5$
communicate with each other.\hfill$\square$

\subsection*{A.5 Completion of the Proof}
The statements (i) and (ii) in Theorem \ref{thm:class} have been proved for the cases of line and cycle graphs. We are now in a place to prove (iii) and (iv) based on our results for tree graphs.
According to Proposition \ref{prop:tree}, for tree graphs without being a line graph, there are five communication classes $\mathbf{J}_n^1$, $\mathbf{J}_n^2$, $\mathbf{J}_n^3$, $\mathbf{J}_n^4$ and $\mathbf{J}_n^5$. Since any connected graph contains a spanning tree, the communication classes of $\mathcal{M}_\mathrm{G}(\mathsf{C}_{\rm pst})$ for any connected graph $\mathrm{G}$ that is not a line or cycle, can only be unions of the $\mathbf{J}_n^j$, $j=1,\dots, 5$.

\medskip

Proof of Theorem \ref{thm:class}(iii):
Suppose $\mathrm{G}$ is neither a line graph nor a cycle graph and it contains no odd cycle. There is a spanning tree of $\mathrm{G}$, denoted $\mathrm{G}_{\mathrm{T}}$. For $\mathcal{M}_{\mathrm{G}_{\mathrm{T}}}(\mathsf{C}_{\rm pst})$,  $\mathbf{J}_n^1$, $\mathbf{J}_n^2$, $\mathbf{J}_n^3$, $\mathbf{J}_n^4$ and $\mathbf{J}_n^5$ are communication classes. $\mathbf{J}_n^1$ and $\mathbf{J}_n^2$ are absorbing states in  $\mathrm{G}$. Because there is no odd cycle, $\mathbf{J}_n^3$ and $\mathbf{J}_n^4$ are the states that any pair of nodes associated with a common edge $\mathrm{G}$ share different values. That is to say, $\mathbf{J}_n^3$ and $\mathbf{J}_n^4$ cannot be accessible from any other states in $\mathcal{M}_\mathrm{G}(\mathsf{C}_{\rm pst})$. Thus, $\mathbf{J}_n^1$, $\mathbf{J}_n^2$, $\mathbf{J}_n^3$, $\mathbf{J}_n^4$ and $\mathbf{J}_n^5$ are still communication classes in $\mathcal{M}_\mathrm{G}(\mathsf{C}_{\rm pst})$, i.e. $\chi_{_{\mathsf{C}_{\rm pst}}}(\mathrm{G})=5$.

\medskip

Proof of Theorem \ref{thm:class}(iv):
Now suppose $\mathrm{G}$ contains an odd cycle. Again, there is a spanning tree $\mathrm{G}_{\mathrm{T}}$ of $\mathrm{G}$. $\mathbf{J}_n^1$, $\mathbf{J}_n^2$, $\mathbf{J}_n^3$, $\mathbf{J}_n^4$ and $\mathbf{J}_n^5$ are communication classes in $\mathcal{M}_{\mathrm{G}_{\mathrm{T}}}( \mathsf{C})$. Also, $\mathbf{J}_n^1$ and $\mathbf{J}_n^2$ are absorbing states in  $\mathcal{M}_\mathrm{G}(\mathsf{C}_{\rm pst})$. For states in $\mathbf{J}_n^3$ and $\mathbf{J}_n^4$, there is an edge $e^*$ belonging to the odd cycle such that the pair nodes of this edge take different values. Now, by choosing another spanning tree $\mathrm{G}_{\mathrm{T}}^*$ containing the edge $e^*$, we can prove that elements in $\mathbf{J}_n^3$, $\mathbf{J}_n^4$ and $\mathbf{J}_n^5$ communicate with each other in $\mathcal{M}_\mathrm{G}(\mathsf{C}_{\rm pst})$. In turn, $\chi_{_{\mathsf{C}_{\rm pst}}}(\mathrm{G})=3$.

\end{document}